\documentclass[a4paper,superscriptaddress,twocolumn,showkeys,pre,aps]{revtex4-2}

\usepackage{latexsym,amsmath}
\usepackage{amsbsy}
\usepackage[pdftex]{graphicx}
\usepackage{amssymb}
\usepackage{epstopdf}
\usepackage[svgnames,dvipsnames,x11names]{xcolor}
\usepackage[colorlinks=true,allcolors=blue]{hyperref}
\usepackage{xspace}

\newcommand{\ZZZ}[2]{}

\renewcommand{\leq}{\leqslant}

\newcommand{\blue}[1]{{\color{blue} #1}}

\newcommand{\D}{\mathcal{D}}
\newcommand{\B}{\mathcal{B}}
\renewcommand{\O}{\mathcal{O}}

\newcommand{\da}{d'Alembertian\xspace}
\renewcommand{\vec}[1]{\mathbf{#1}}

\begin{document}

    \title{Local d'Alembertian for causal sets}
	\date{\today}
	\author{Mari\'an Bogu\~n\'a}
	\email[]{marian.boguna@ub.edu}
	\affiliation{Departament de F\'isica de la Mat\`eria Condensada, Universitat de Barcelona, Mart\'i i Franqu\`es 1, E-08028 Barcelona, Spain}
	\affiliation{Universitat de Barcelona Institute of Complex Systems (UBICS), Universitat de Barcelona, Barcelona, Spain}
	
	\author{Dmitri Krioukov}
	\email[]{dima@northeastern.edu}
	\affiliation{Department of Physics \& Network Science Institute \&\\ Department of Mathematics \& Department of Electrical and Computer Engineering, Northeastern University, Boston, Massachusetts, USA}
			
\begin{abstract}

Causal set theory is an intrinsically nonlocal approach to quantum gravity, inheriting its nonlocality from Lorentzian nonlocality. This nonlocality causes problems in defining differential operators---such as the d'Alembert operator, a cornerstone of any relativistic field theory---within the causal set framework. It has been proposed that d'Alembertians in causal sets must themselves be nonlocal to respect causal set nonlocality. However, we show that such nonlocal d'Alembertians do not converge to the standard continuum d'Alembertian for some basic fields. To address this problem, we introduce a local d'Alembert operator for causal sets and demonstrate its convergence to its continuum counterpart for arbitrary fields in Minkowski spacetimes. Our construction leverages recent developments in measuring both timelike and spacelike distances in causal sets using only their intrinsic structure. This approach thus reconciles locality with Lorentz invariance, and paves a way toward defining converging discrete approximations to locality-based differential operators even in theories that are inherently nonlocal.

\end{abstract}
\maketitle

%\tableofcontents

\section{Introduction}
\label{sec:intro}

Causal set theory (CST)~\cite{Surya:2019aa} is a peculiar actor in the neverending dramatic quest for a theory of quantum gravity. CST is one of the most minimalistic approaches there, in terms of the assumptions it makes. It simply posits that spacetime, at the Planck scale, is not a continuous manifold but rather a discrete object made up of elementary spacetime atoms, with no internal structure, connected by causal  relations~\cite{Bombelli:1987im,bombelli1988,meyer1989,sorkin1991,sorkin1997,rideout1999,sorkin2003,sorkin2005,henson2006,dowker2006,surya2008,henson2010,surya2012}. The approach finds its origins in the work of Hawking, King, and McCarthy~\cite{hawking_king_mccarthy_1976}, as well as Malament~\cite{malament1977}, who demonstrated that spacetimes that satisfy some natural niceness assumptions, and that have the same causal structure, are equivalent modulo conformal rescaling.

CST suggests that the smooth continuum of spacetime that we observe at macroscopic scales is an illusion, a coarse-grained smearing of an underlying fine-grained discrete structure at the Planck scale. The process by which this discrete structure transitions to the smooth spacetime continuum is known as the continuum limit of causal sets~\cite{major2007recovering,brightwell20082d,surya2012evidence,saravani2014causal,belenchia2016continuum,machet2020continuum}. Achieving this limit requires an ability to measure distances using only the causal set structure, which is relatively straightforward for time-like separated events~\cite{bombelli1988}, but considerably more challenging for space-like separated events~\cite{Brightwell:1991aa,Rideout_2009,Eichhorn_2019,Boguna:2024qf}.

CST has traditionally been viewed as a nonlocal theory~\cite{sorkin2009does}, a perspective that arises from the observation that in causal sets obtained from random sprinklings on Lorentzian manifolds, the number of nearest neighbors of a given event is infinite and distributed on a hyperbolic surface of constant proper time of the order of the Planck time. These nearest neighbors, although separated by one unit of proper time, can be arbitrarily distant in space. Yet thanks to Lorentz invariance, there is seemingly nothing that can differentiate one such neighbor from another, so all of them must be considered on the same footing in any field theory defined on the causal set.

This line of reasoning inspired the development of nonlocal d'Alembert operators for causal sets~\cite{sorkin2009does,Benincasa:2010eu,Dowker_2013,Glaser_2014,Aslanbeigi:2014sy,dowker2024timelikeboundarycornerterms}. These operators are nonlocal in that they aim to account equally for all events in the aforementioned infinite set of the nearest neighbors of a given event in a causal set. Yet at the same time they attempt to resurrect locality in the continuum limit attained at the infinite event density. While these two goals appear mutually contradictory, nonlocal discrete d'Alembertians have nevertheless succeeded in converging to the continuum limit when acting upon many types of scalar fields. Many but not all. Unlike their continuous parent, the standard d'Alembertian, these nonlocal operators lead to divergences when acting on certain scalar fields, including the simplest one, a constant field, as we show in Section~\ref{sec:nonlocal}. These divergences can be suppressed by requiring that the scalar field upon which the operator acts is zero everywhere except a bounded chunk of spacetime. However, this requirement is equivalent to predetermining the set of events used in the evaluation of the operator, which amounts to defining a specific reference frame. More importantly, it is reasonable to require a discrete d'Alembertian to converge to the continuous one on any good-looking field, not only on some specific fields.

The key idea behind our work presented here is very simple. While all neighbors of a given event in a causal set are indeed equivalent with respect to this event, they are not equivalent among themselves. In~\cite{Boguna:2024qf}, we have shown that using only the the causal set structure, it is possible to measure spatial distances between causal set events all the way down to the smallest possible distances of the order of the Planck length. This means that we can take one of those neighboring events, and then sort the rest of them in the order of increasing proper length from the selected neighbor. This implies that local neighborhoods can be defined and, consequently, local differential operators can be constructed without breaking Lorentz invariance.

Our main contributions in this paper are a definition of a local discrete d'Alembert operator based on the idea above, and an analytic and numerical evidence that this operator converges to the continuous d'Alembertian on any scalar field in the $(d+1)$-dimensional Minkowski spacetime.

We begin with an illustration of key components of our construction in much simpler settings in Section~\ref{sec:laplacian}, where we define a discrete field Laplacian for random geometric graphs in the Euclidean plane, and show that this Laplacian converges to the standard continuous Laplacian acting on the same field. We then proceed to Section~\ref{sec:nonlocal} where we discuss key complications arising from Lorentzian nonlocality, including the divergences of nonlocal discrete d'Alembertians mentioned above. Finally, by far the longest Section~\ref{sec:Alembertian}, with its own internal structure, contains a detailed account of our discrete d'Alembertian construction, and the analytical and numerical demonstrations of its convergence to the continuous d'Alembert operator in the continuum limit. Some concluding remarks are in Section~\ref{sec:conclusions}.

\begin{figure}[t]
\centerline{\includegraphics[width=0.8\columnwidth]{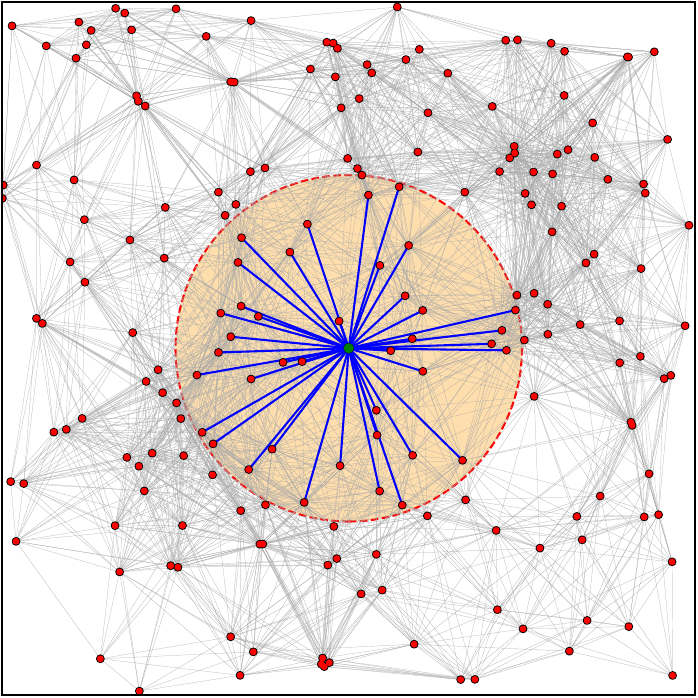}}
\caption{{\bf Random geometric graphs.} The figure shows a random geometric graph generated by the Poisson point process on the Euclidean unit square with point density $\rho = 200$, and connection radius $r_c = 0.26$. The blue links within the shaded circle represent the neighbors of the central node in green. The discrete Laplacian in Eq.~\eqref{eq:laplace3}, evaluated at the central node, is computed using the values of the scalar field at these neighboring nodes.
}
\label{fig:grg}
\end{figure}

\section{Laplacian in random geometric graphs}
\label{sec:laplacian}

To illustrate the key ideas behind our discrete \da and its convergence to the continuum one, we first consider the Riemannian case, which is simpler than the Lorentzian case, yet all the main ingredients are present, except one---nonlocal connections. In Riemannian manifolds, the \da is the Laplacian, while the role of causal sets is played by random geometric graphs. In these graphs, only those node pairs that are close to each other in the space are connected. In this section, we illustrate how the discrete structure of these only-local connections can be used to evaluate the action of the Laplacian on a scalar field. This example provides a clear introduction to the techniques that we will later apply to the more complex Lorentzian settings that involve nonlocal connections. To the best of our knowledge, the results in~\cite{BELKIN20081289}, dealing with the convergence of a discrete Laplacian defined for point clouds in Riemannian manifolds, are closest in spirit to the results presented in this section; there are also many results concerning the convergence of the standard graph Laplacian of random geometric graphs~\cite{Hein:2007aa,Hamidouche:2023uq}, as opposed to the field Laplacian we are considering here.

Consider a random geometric graph in the Euclidean plane, generated by a Poisson point process with density $\rho$, and a connection radius $r_c$. In this graph, two nodes are connected if and only if the Euclidean distance between them is less than $r_c$.
The degrees of nodes are Poisson-distributed random variables with mean $\langle k \rangle=\rho \pi r_c^2$ and variance $\sigma_k^2=\langle k \rangle$. Therefore, in the limit $\langle k \rangle \gg 1$, the relative fluctuations $\sigma_k/\langle k \rangle$ vanish, so we can assume that all nodes have approximately the same number of neighbors $\langle k \rangle$. Figure~\ref{fig:grg} shows a visualization of a random geometric graph, highlighting a central node and its neighbors. 

Since the graph can be thought of as a random discretization of the Euclidean plane and distances in it, we are tempted to ask to what extent this discrete sampling of a continuous manifold can recover its continuous properties. Specifically, we are interested in evaluating the Laplace operator acting on a scalar field $\phi(x,y)$ defined at every point on the plane, and, consequently, at any node of the graph. If we evaluate the Laplacian using only the graph structure and the value of the $\phi$-field at its nodes, will it converge to the Laplacian evaluated on the plane in a certain limit? 

The answer to this question is positive, and to illustrate this convergence, let us evaluate the graph Laplacian at a particular node, which we assume, without loss of generality, is located at the origin, so we call this node~$0$. In the limit $r_c \rightarrow 0$, the scalar field at $0$'s neighbor $i$ at coordinates $\vec{r}_i=(x_i,y_i)$ can be approximated using the Taylor series as
\begin{widetext}
\begin{equation}
\phi(x_i,y_i) \equiv \phi_i = \phi_0 + \nabla \phi_0 \cdot \vec{r}_i + \frac{1}{2} \left( \left. \frac{\partial^2 \phi}{\partial x^2}\right|_0 x_i^2 + \left. \frac{\partial^2 \phi}{\partial y^2}\right|_0 y_i^2 + 2 \left. \frac{\partial^2 \phi}{\partial x \partial y}\right|_0 x_i y_i \right) + \mathcal{O} \left( r_c^3 \right),
\quad x_i,y_i < r_c,
\label{eq:laplace1}
\end{equation}
\end{widetext}
where the subscript $0$ indicates that the field and its derivatives are evaluated at the node $0$, located at the origin $x = y = 0$. We now want to take the average of Eq.~\eqref{eq:laplace1} over the set of $k_0$ neighbors of the 0 node. For that end, we define their average $x$-coordinates (squared) as
\begin{equation}
\bar{x} \equiv \frac{1}{k_0} \sum_{i=1}^{k_0} x_i, \quad \bar{x^2} \equiv \frac{1}{k_0} \sum_{i=1}^{k_0} x_i^2,
\end{equation}
and similarly for the $y$-coordinates. The quantities $\bar{x}$ and $\bar{x^2}$ are still random variables since they depend on a particular realization of the Poisson point process. Due to the Euclidean symmetries, the mean of $\bar{x}$ is zero, $\langle \bar{x} \rangle = 0$, while its variance is $\langle x^2 \rangle / \langle k \rangle$, where $\langle {x^2} \rangle \equiv \langle \bar{x^2} \rangle$ is the mean of $\bar{x^2}$. One can check that $\langle x^2 \rangle = r_c^2/4$, substituting which into the average of~\eqref{eq:laplace1}, we observe that the average $\bar{\phi}$ of the values of the scalar field $\phi$ over the neighbors of node $0$ can be written as
\begin{equation}
\bar{\phi} = \phi_0 + \frac{1}{2} \nabla^2 \phi_0 \langle x^2 \rangle + \mathcal{O} \left(\sqrt{\frac{\langle x^2 \rangle}{\langle k \rangle}} \right) + \mathcal{O} \left( r_c^4 \right),
\end{equation}
where the first error term comes from $\bar{x}$, and the second from the truncated Taylor expansion up to fourth order. Using $\langle x^2 \rangle = r_c^2/4$ and $\langle k \rangle=\rho \pi r_c^2$, we can then write the Laplacian as
\begin{equation}
\nabla^2 \phi_0 = \frac{8}{r_c^2}(\bar{\phi} - \phi_0) + \mathcal{O} \left(\frac{1}{r_c^2 \sqrt{\rho}} \right) + \mathcal{O} \left( r_c^2 \right).
\label{eq:laplace2}
\end{equation}

The last equation says that the continuous Laplacian $\nabla^2 \phi_0$ of the field evaluated at the origin can be recovered in the $r_c\to0$ limit from the discrete average $\bar{\phi}$ of the field over the neighbors of the $0$ node in the random geometric graph.
That is,
\begin{equation}
\mathcal{D} \phi_0 = \frac{8}{r_c^2}(\bar{\phi} - \phi_0)
\label{eq:laplace3}
\end{equation}
can be understood as a discrete Laplacian associated with our random geometric graph. Yet this continuous recovery is possible, and the discrete Laplacian converges to its continuous father,
\begin{equation}
\lim_{r_c\to0}\mathcal{D} \phi_0 = \nabla^2 \phi_0,
\end{equation}
only if the both error terms in~\eqref{eq:laplace2} are small in the limit. They are both small if
\begin{equation}
\frac{1}{\rho^{1/4}} \ll r_c \ll 1.
\end{equation}
This condition is fulfilled in the continuum limit $\rho\to\infty$ if $r_c$ is chosen to be $r_c \sim \rho^{-\delta}$ with $0 < \delta < 1/4$.
This limit means that we are sampling the space more and more densely ($\rho\to\infty$), while links in our random geometric graph become increasingly microscopic ($r_c\to0$), thus approximating distances in the space with growing precision. There is also the optimal value of $\delta = 1/8$ making the scaling of both error terms in~\eqref{eq:laplace2} with~$\rho$ the same~$\mathcal{O}\left( \rho^{-1/4} \right)$, so Eq.~\eqref{eq:laplace2} becomes
\begin{equation}\label{eq:laplace-final}
  \nabla^2 \phi_0 = \mathcal{D} \phi_0 + \mathcal{O}\left( \rho^{-1/4} \right).
\end{equation}

\begin{figure}[t]
\centerline{\includegraphics[width=\columnwidth]{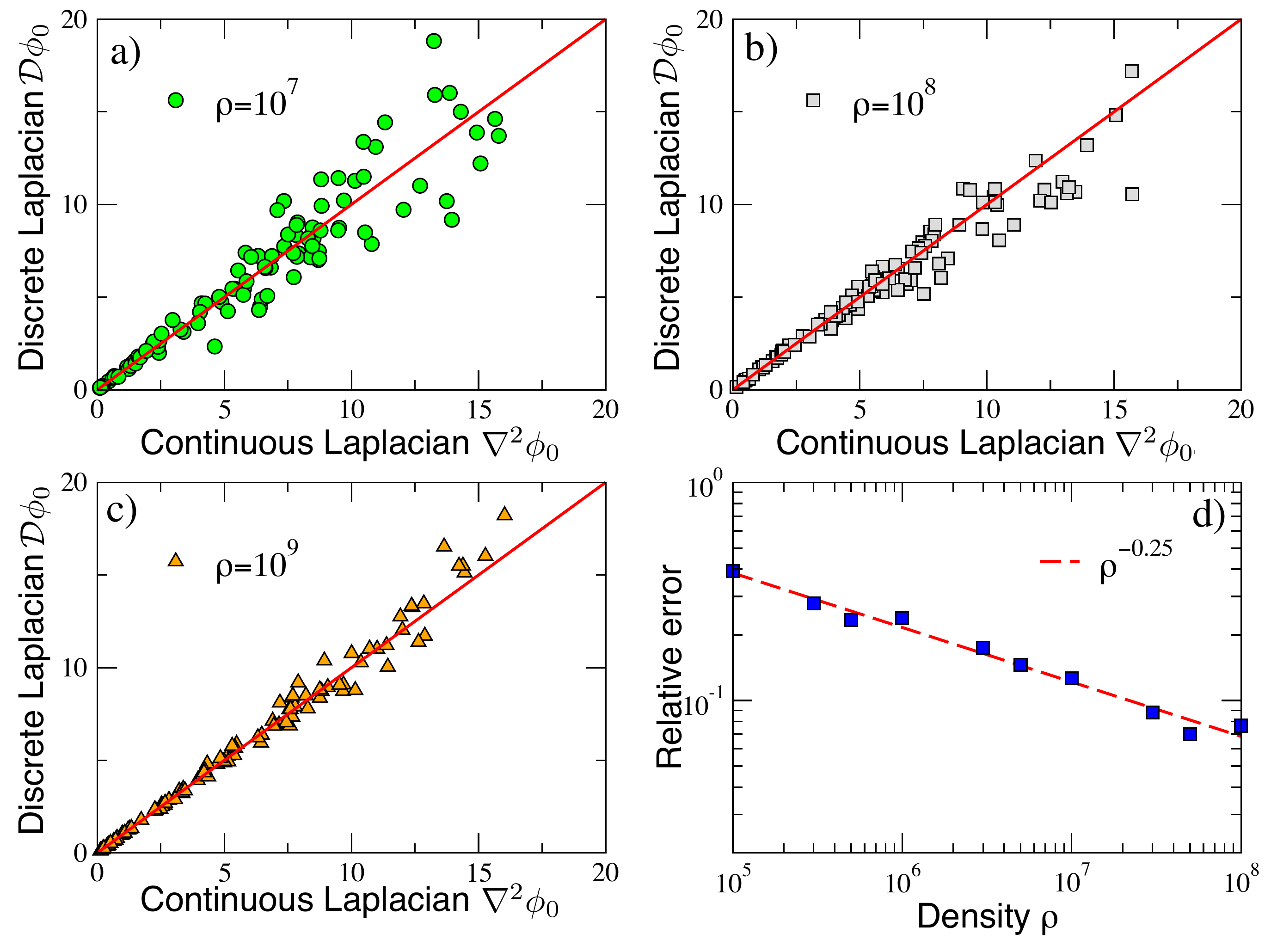}}
\caption{{\bf Laplacian simulations.} \textbf{(a-c)} The values of the discrete Laplacian $\mathcal{D}\phi_0$ in Eq.~\eqref{eq:laplace3} acting on the field $\phi=e^{ax+by}$ at $x=y=0$ vs.\ the continuous value $\nabla^2 \phi_0=a^2+b^2$ for different point densities~$\rho$. For each value of $a$ and $b$, a separate realization of the Poisson point process within the disk $r<r_c = 0.75 \rho^{-1/8}/\sqrt{a^2+b^2}$ is sampled. \textbf{(d)} The average relative error of the discrete vs.\ continuous Laplacian, defined as $\left\langle\left|\left(\mathcal{D}\phi_0/\nabla^2 \phi_0\right) - 1\right|\right\rangle$, for different values of $\rho$. For each value of~$\rho$, the relative error is averaged of 100 random values of $a,b$ and Poisson point process realizations. The red dashed line is the predicted scaling behavior $\rho^{-1/4}$ from Eq.~\eqref{eq:laplace-final}.
}
\label{fig:laplacian}
\end{figure}

To validate this result in simulations, we choose the scalar field $\phi(x,y) = e^{ax+by}$, for which $\nabla^2 \phi_0 = a^2 + b^2$, and then sample 100 values of $a$ and $b$ selected uniformly at random within the interval $[0,3]$. For each pair of values $(a,b)$, we generate random geometric graphs at different densities $\rho$, with the connection radius scaling as $r_c \sim \rho^{-1/8}$. We then compute the discrete Laplacian in the generated graphs according to Eq.~\eqref{eq:laplace3}, and compare the results of these computations with the continuous value $a^2+b^2$. The outcomes of these experiments shown in Fig.~\ref{fig:laplacian} confirm the expected: the discrete Laplacian converges to the continuous one as $\rho$ increases.

These results provide a simple illustration that differential operators on smooth manifolds can be approximated using discrete random samplings of the manifold, and that under appropriate scaling conditions, such approximations can become exact in the continuum limit when the sampling density approaches infinity.

The key idea behind the approximation used in the discrete Laplacian in Eq.~\eqref{eq:laplace3} is fairly straightforward. Indeed, since the Laplacian operator is a measure of how fast the field changes around a point, the properly rescaled difference of the values of the field at a node and its neighbors in a random geometric graph converges to the continuous Laplacian. 

However, while in Riemannian geometry, this convergence is relatively easy to prove, the intrinsic nonlocality of Lorentzian geometry complicates things significantly. We discuss these complications next. 

\section{Nonlocal d'Alembertians in causal sets}\label{sec:nonlocal}

In this section we show how Lorentzian nonlocality causes problems---divergencies---in discrete nonlocal d'Alembertians applied to nonlocal fields.

Causal sets built on top of Lorentzian manifolds are akin to random geometric graphs in Riemannian manifolds. The vertices in both cases are realizations of Poisson point processes on a manifold, but edges in causal sets reflect causality instead of spatial proximity: two vertices in a causal set are linked if they are timelike-separated in the spacetime manifold. We will refer to vertices in causal sets as \emph{events}, and will assume that links implied by transitivity are removed.

The Laplace operator in Lorentzian manifold is the d'Alembert operator, which in natural units is $\Box \equiv \eta^{\mu \nu} \partial_\mu \partial_\nu = -\partial_t^2+\nabla^2$ in Minkowski spacetime. While this operator is perfectly local when defined in continuous spacetime, its satisfactory definition in a causal set that discretely samples this spacetime is not immediately clear. Indeed, if Minkowski spacetime is sampled at finite density $\rho$, each event in the sampling is directly connected to an infinite number of other events lying approximately on the hyperboloid of constant proper time $\tau_c \propto \rho^{-1/(d+1)}=t_P$, where $d$ is the dimension of the spatial part of the metric tensor, and $t_P$ is defined to be the Planck time. Therefore, while these neighbors of the event are at a small temporal distance from it, they can be arbitrarily far apart among themselves in terms of space-like distance. This observation suggests that discrete versions of Lorentzian manifolds are intrinsically nonlocal, and thus, operators acting on them should also be nonlocal.

Nonlocal d'Alembertians acting on a scalar field $\phi$ at a given event $a_0$ were first introduced in~\cite{sorkin2009does} for the $\mathbb{M}^{1+1}$ Minkowski spacetime, extended later to $\mathbb{M}^{3+1}$ in~\cite{Benincasa:2010eu}, and generalized further in a number of papers~\cite{Dowker_2013,Glaser_2014,Aslanbeigi:2014sy,dowker2024timelikeboundarycornerterms}, where they appear as a basis for a definition of action in causal sets. The main goal behind such definitions is to come up with discrete d'Alembertians that are linear combinations of the scalar field evaluated at specific events in the past of~$a_0$, and that satisfy the following key requirements~\cite{Aslanbeigi:2014sy}:
\begin{enumerate}
\item \textbf{invariance:} a discrete d'Alembertian $B$ must be Lorentz invariant,
\item \textbf{convergence:} the expectation of $B$'s values averaged over Poisson sprinklings of increasing densities~$\rho$ must converge to the continuous d'Alembertian~$\Box$ in the continuum limit where the sprinkling density goes to infinity:
    \begin{equation}
      \lim_{\rho\to\infty}\langle B_\rho \rangle = \Box.
    \end{equation} 
\end{enumerate}
The existing $B$-definitions satisfy all these requirements, although to satisfy the last one---the convergence in the continuum limit---they require that the field $\phi$ to which $B$ is applied must have a compact support. If applied to a noncompact field, $B$ may not converge.

To illustrate this point, let us consider the two-dimensional Minkowski spacetime for simplicity; similar results hold in other dimensions. In two dimensions, the simplest nonlocal d'Alembertian operator $B_\rho$ acting on the scalar field $\phi$ at $a_0$ is defined as~\cite{sorkin2009does}
\begin{widetext}
\begin{equation}
\left(B_\rho^{(2)} \phi \right)(a_0) \equiv \rho \left[ -2 \phi(a_0) + 4 \left( \sum_{{\bf x} \in I_0(a_0)} \phi({\bf x}) - 2\sum_{{\bf x} \in I_1(a_0)} \phi({\bf x}) + \sum_{{\bf x} \in I_2(a_0)} \phi({\bf x}) \right) \right],
\label{eq:1}
\end{equation}
\end{widetext}
where $I_n(a_0)$ is the set of events ${\bf x}$ in the past of $a_0$ such that the number of events in the Alexandrov set between $a_0$ and ${\bf x}$ is $n$. Given the symmetry of Minkowski spacetime, the same definition can be made using events in the future of $a_0$ instead of its past.

Equation~\eqref{eq:1} defines a random variable that depends on a particular realization of the Poisson point process. The average of this random variable over Poisson point process realizations is given by
\begin{widetext}
\begin{equation}
\left\langle \left(B_\rho^{(2)} \phi \right)(a_0) \right\rangle = \rho \left[ -2 \phi(a_0) + 4\rho \left( \int_{{\bf x} \in \mbox{Past}(a_0)} \sqrt{-g}\,d{\bf x}\,e^{-\rho V({\bf x}|a_0)} \left(1 - 2 \rho V({\bf x}|a_0) + \frac{1}{2} \left(\rho V({\bf x}|a_0)\right)^2 \right) \phi({\bf x}) \right) \right]
\label{eq:2}
\end{equation}
where the integral extends over the past lightcone of $a_0$, and $V({\bf x}|a_0)$ is the volume of the Alexandrov interval between $a_0$ and ${\bf x}$. We next rewrite Eq.~\eqref{eq:2} as
\begin{equation}
\left\langle \left(B_\rho^{(2)} \phi \right)(a_0) \right\rangle = \rho \left[ -2 \phi(a_0) + 4\rho \left(1 + 2 \rho \frac{d}{d\rho} + \frac{1}{2}\rho^2 \frac{d^2}{d \rho^2} \right)I(\rho) \right]
\label{eq:3}
\end{equation}
\end{widetext}
where
\begin{equation}
I(\rho) = \int_{{\bf x} \in \mbox{Past}(a_0)} \sqrt{-g}\,d{\bf x}\,e^{-\rho V({\bf x}|a_0)} \phi({\bf x}).
\end{equation}
Next, without loss of generality, we set $a_0$ to be the origin, $a_0=(0,0)$, and thanks to the symmetry of Minkowski spacetime, we use the future of $a_0$ instead of its past. The future lightcone $(t,x)$ of $a_0$ can be foliated by hyperbolas with hyperbolic coordinates~$(\tau,\chi)$ given by
\begin{align}
  t &= \tau \cosh{\chi},\label{eq:future-t} \\
  x &= \tau \sinh{\chi},\label{eq:future-x}
\end{align}
where $\tau \in (0,\infty)$ is the proper time from $a_0$ to the hyperbola, and $\chi \in (-\infty,\infty)$ is its spatial coordinate.
Let us now define the dimensionless proper time
\begin{equation}
  \hat{\tau} = \sqrt{\frac{\rho}{2}} \tau,
\end{equation}
and the integral
\begin{equation}
\hat{I}(\rho,\chi) \equiv \frac{2}{\rho} \int_{0}^{\infty} d\hat{\tau}\,\hat{\tau}\,e^{-\hat{\tau}^2} \phi \left( \sqrt{\frac{2}{\rho}} \hat{\tau} \cosh{\chi}, \sqrt{\frac{2}{\rho}} \hat{\tau} \sinh{\chi} \right).
\label{eq:boxaverage}
\end{equation}
With these notations, the integral $I(\rho)$ in Eq.~\eqref{eq:3} becomes $I(\rho)=\int_{-\infty}^{\infty} I(\rho,\chi)\,d\chi$, so Eq.~\eqref{eq:3} can be rewritten as
\begin{widetext}
\begin{equation}
\left\langle \left(B_\rho^{(2)} \phi \right)(a_0) \right\rangle = \rho \left[ -2 \phi(a_0) + 4\rho \int_{-\infty}^{\infty} d \chi \left(1 + 2 \rho \frac{d}{d\rho} + \frac{1}{2}\rho^2 \frac{d^2}{d \rho^2} \right)\hat{I}(\rho,\chi) \right],
\label{eq:4}
\end{equation}
\end{widetext}
We next use this expression to analyze the d'Alembertian operator $B_\rho^{(2)}$ acting upon simple test fields.

\textbf{Example 1: constant scalar field.}
If the scalar field $\phi$ is constant, $\phi(t,x)=\phi_0$, then the continuous d'Alembertian applied to this filed yields zero,
\begin{equation}
  \Box \phi = 0.
\end{equation}
To see what we get from the discrete d'Alembertian, observe that $\hat{I}(\rho,\chi)=\phi_0/\rho$, substituting which into Eq.~\eqref{eq:4}, we get
\begin{equation}
\left(1 + 2 \rho \frac{d}{d\rho} + \frac{1}{2}\rho^2 \frac{d^2}{d \rho^2} \right) \hat{I}(\rho,\chi) = 0,
\end{equation}
leading to
\begin{equation}
\left\langle \left(B_\rho^{(2)} \phi \right) (a_0) \right\rangle = -2 \rho \phi_0.
\label{eq:5}
\end{equation}
That is, instead of being zero, the discrete nonlocal d'Alembertian diverges in the limit $\rho \rightarrow \infty$, unless $\phi_0=0$. A similar divergence takes place with any scalar field which either converges to a constant in the limit $x \rightarrow \pm \infty$, or is a growing function of $x$, or does not go to $0$ at large~$x$ sufficiently fast.

\textbf{Example 2: scalar field as a function of $\tau^2=t^2 - x^2$.}
As an example of an even worse divergence, let us consider the scalar field given by
\begin{equation}
\phi(t,x) = \tilde\phi(\tau^2)=e^{-\tau^2},
\label{eq:phiexample2}
\end{equation}
for which the continuous d'Alembertian yields
\begin{equation}
  \Box \phi(0,0) = 4.
\end{equation}
Turning to the discrete d'Alembertian, Eq.~\eqref{eq:boxaverage} yields
\begin{equation}
\hat{I}(\rho,\chi) = \frac{1}{2 + \rho},
\end{equation}
which does not depend on $\chi$, so Eq.~\eqref{eq:4} says that
\begin{equation}
\left\langle \left(B_\rho^{(2)} \phi \right) (a_0) \right\rangle = -2 \rho + \frac{16 \rho^2}{(2 + \rho)^3} \int_{-\infty}^{\infty} d \chi = \infty.
\label{eq:6}
\end{equation}
That is, in this case, the discrete d'Alembertian diverges even for any finite $\rho$ due to the divergence of the integral over $\chi$.
A similar divergence holds for any field which is a function of $\tau^2$.

The observed lack of the convergence of the discrete nonlocal d'Alembertian to the continuous one in the examples above arises from the fact that the considered test fields have noncompact supports. The convergence is restored if the field is set to zero outside of a finite-length interval.

To see this explicitly, let us consider the constant field $\phi(t,x)=\phi_0$ for $x \in [-L, L]$, while at $|x| > L$ the field is set to $0$. In this case, the integral $I(\rho)$ becomes
\begin{equation}
I(\rho) = \frac{2\sqrt{\pi}\phi_0}{\rho} \int_0^{L\sqrt{\frac{\rho}{2}}} e^{x^2} \text{Erfc}(x) \, dx,
\end{equation}
where $\text{Erfc}(x)$ is the complementary error function. Plugging this expression into Eq.~\eqref{eq:3}, we get
\begin{equation}
\left\langle \left(B_\rho^{(2)} \phi \right) (a_0) \right\rangle = \rho \phi_0 \left[ -2 + \sqrt{\pi}s(3 + 2s^2)e^{s^2}\text{Erfc}(s) - 2s^2 \right],
\end{equation}
where $s = L \sqrt{\rho/2}$. In the limit $\rho \rightarrow \infty$, this expression is asymptotically
\begin{equation}\label{eq:L-finite}
\left\langle \left(B_\rho^{(2)} \phi \right) (a_0) \right\rangle \sim \frac{1}{L^4 \rho}.
\end{equation}
Therefore, for any fixed $L$, the discrete nonlocal d'Alembertian converges to zero, as it should on a constant field.

This result holds for an interval of any size $L$ as long as it is finite. In view of this observation, one may hope that the limit $\rho \rightarrow \infty$ could lead to the correct continuum limit for fields with noncompact supports. However, even if $L$ is taken to be arbitrarily large, the contribution to the nonlocal d'Alembertian from the region $|x| > L$ remains non-negligible. Consequently, the order of limits matter: if we first take $\rho \rightarrow \infty$ and then $L \rightarrow \infty$, then we get $0$ according to Eq.~\eqref{eq:L-finite}, but if we first take $L \rightarrow \infty$ and then $\rho \rightarrow \infty$, then we get $\infty$ according to Eq.~\eqref{eq:5}. In the former case, we are dealing with constant fields with compact supports, while in the latter case, our constant field has a noncompact support.

A related important observation, which is typical for situations like here, is that we can always do some mathematical tricks that do not make any physical sense to obtain any desirable outcome. For instance, we can take the limit $\rho\to\infty$ while simultaneously tending $L$ to $\infty$ as some growing function of $\rho$. If we do this, then we can choose this function to be such that $\left\langle \left(B_\rho^{(2)} \phi \right) (a_0) \right\rangle$ converges to any arbitrary value.
Similar observations can be made for fields that depend on $\tau^2$.

These observations show that discrete nonlocal d'Alembert operators converge to the correct continuum limit only on \emph{local} fields---that is, fields with compact supports. This restriction is a form of locality requirement.
In the next section, we construct a \emph{local} d'Alembertian operator that converges to the correct continuum limit at $\rho \rightarrow \infty$ on any field, local or nonlocal.

\section{Local d'Alembertian in causal sets}
\label{sec:Alembertian}

Here we construct a d'Alembert operator for causal sets that converges to the correct continuum limit on any field. We prove this convergence for causal sets sprinkled into Minkowski spacetimes. Our operator is \emph{local} in the sense that it operates on local neighborhoods of an event in a causal set. These neighborhoods are defined in Section~\ref{ssec:neighbor}.

The main challenge behind their definition is that it cannot refer to any information about the continuous manifold onto which the causal set is sprinkled. Indeed, a general causal set is not obtained by sprinkling into any spacetime. The causal sets that we deal with in this section are sprinkled into Minkowski spacetime just to prove the d'Alembertian convergence. But since a causal set may not be (directly) associated with any continuous spacetime, a causal set d'Alembertian definition may use only information contained in the causal set itself. Therefore, no part of the operator definition, including the definition of our local neighborhoods, is allowed to use any manifold information either. Fortunately, the definition of our local neighborhoods relies only on proper temporal and spatial distances, both of which can be reliably estimated from the causal set structure alone according to the past results that we review in Section~\ref{ssec:proper}.

With the local neighborhoods defined Section~\ref{ssec:neighbor}, we then show, in Sections~\ref{ssec:temporal} and~\ref{ssec:spatial}, how they can be used to define discrete temporal and spatial derivatives, again using only the causal set information. Combining the two derivatives, we arrive at the definition of our discrete d'Alembertian in Section~\ref{ssec:box}, whose expected deviation from the continuous d'Alembertian, we show, converges to zero in the continuum limit. We confirm these theoretical results in simulations in Section~\ref{ssec:sims}.

For our purposes here, we will treat a general causal set~$\mathcal{C}$ as a directed acyclic graph in which links implied by transitivity are removed. That is, if $x,y\in\mathcal{C}$ and $x \prec y$, meaning that $x$ is in the past of $y$, then there is a directed link from $x$ to $y$ if and only if there is no other element $z \in \mathcal{C}$ such that $x \prec z \prec y$.

\subsection{Measuring proper times and lengths}\label{ssec:proper}

\subsubsection{Temporal distances}\label{ssec:proper-time}

The measurement of temporal distances in causal sets is relatively straightforward. The proper time $\tau_{\cal C}(a,b)$ between $a \prec b$ in $\cal{C}$ is proportional to the length $n_{\mathcal{C}}(a,b)$ of the longest chain from $a$ to $b$, because this chain represents the geodesic between the two events~\cite{Brightwell:1991aa}. The convergence of this discrete proper time definition to its continuum limit has been proven for causal sets that are Poisson-sprinkled into Minkowski spacetimes ${\mathbb M}^{d+1}$.

Specifically, let us define the Planck time~$t_P$ via the sprinkling density~$\rho$ as
\begin{equation}\label{eq:rho-tP}
t_P=\rho^{-1/(d+1)},
\end{equation}
and the discrete proper time as
\begin{equation}\label{eq:tauc}
\tau_{\cal C}(a,b)\equiv \alpha_d t_P n_{\cal C}(a,b),
\end{equation}
where $\alpha_d$ is a constant that depends on the spacetime dimension only. Then it was shown in~\cite{Bollobas:1991aa,Brightwell:1991aa,Bachmat:2008aa} that
\begin{equation}
\tau_{\cal C}(a,b)=\tau_{{\mathbb M}^{d+1}}(a,b) + t_P^{1-\beta_d} \zeta_{d},
\label{error_tau}
\end{equation}
where $\tau_{{\mathbb M}^{d+1}}(a,b)$ is the proper time between $a \prec b$ in ${\mathbb M}^{d+1}$, $0\leq\beta_d<1$, and $\zeta_d$ is a random variable with bounded fluctuations and negative average. The values of constants $\alpha_d$ and $\beta_d$ are known exactly only for $d=1$. They are $\alpha_1=1/\sqrt{2}$~\cite{LOGAN1977206} and $\beta_1=1/3$~\cite{Baik1999}. In higher dimensions, the approximate values of $\beta_d$ are known to be $\beta_2 \approx 1/4$, $\beta_3 \approx 1/6$, and $\beta_d \approx 0$ for $d \ge 4$~\cite{Lssig1998OnGD}, while our numeric experiments in Appendix~\ref{sec:alpha2} suggest that $\alpha_2=1/\sqrt{2}$ as well. Even though the exact values of $\beta_d$ are not known for $d>1$, the fact that $\beta_d<1$ in Eq.~\eqref{error_tau} implies that
\begin{equation}
  \lim_{\rho \rightarrow \infty}\tau_{\cal C}(a,b) = \tau_{{\mathbb M}^{d+1}}(a,b),
\end{equation}
so that proper times can be recovered with arbitrary precision simply by counting links in ${\mathcal{C}}$.

\subsubsection{Spatial distances}

The measurement of spatial distances in causal sets is much more challenging, yet the problem has been recently solved in~\cite{Boguna:2024qf}. Proper lengths between pairs of unrelated events $a$ and $b$ are defined there to be some functions of the causal overlap between $a$ and $b$, which is the intersection of the past lightcones of the two events with the future lightcone of another event $c$ in the common past of $a$ and $b$. This discrete spatial distance definition was shown to converge to the correct continuum limit in~\cite{Boguna:2024qf} for causal sets that are Poisson-sprinkled into ${\mathbb M}^{d+1}$, even for pairs of events whose spatial separation is of the order of the Planck length. 

\subsection{Defining local neighborhoods}\label{ssec:neighbor}

\subsubsection{Temporal neighborhoods}\label{ssec:neighbor-temporal}

We define our temporal neighborhoods based on inertial reference frames in $\mathcal{C}$, which are geodesic time-like curves, with one of the events in the geodesic chosen as the origin.

\begin{figure}[t]
\centerline{\includegraphics[width=\columnwidth]{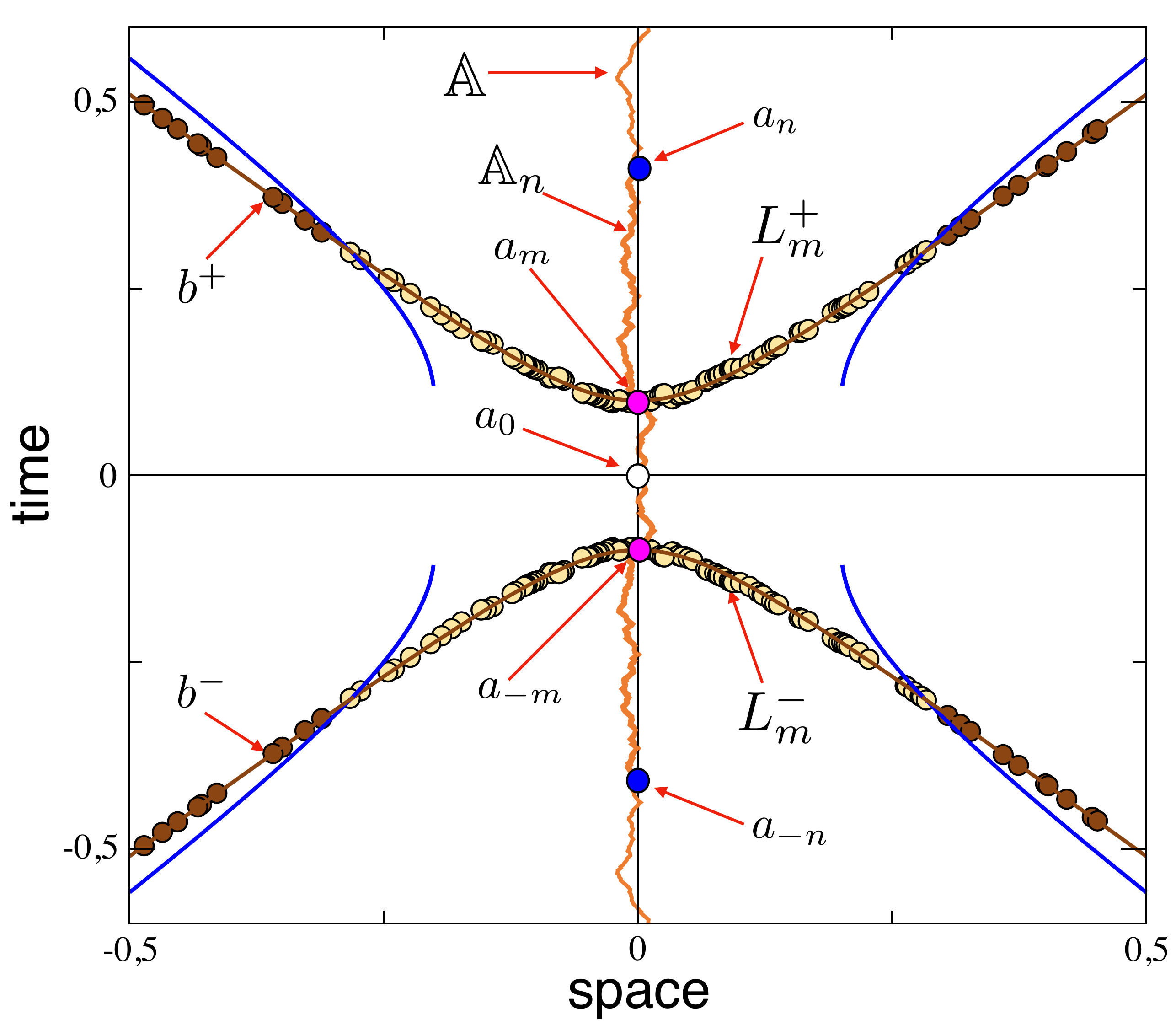}}
\caption{{\bf Defining local neighborhoods.} $100,000$~events (not shown) are sprinkled in the $[-0.5,+0.5]^2$ box in $\mathbb{M}^2$ to illustrate the key elements of the local discrete d'Alembertian construction in Section~\ref{sec:Alembertian}. In this sprinkling, the point density is thus $\rho=10^5$\ and $t_P=3.16\times10^{-3}$. Event $a_0$ is at the origin; geodesic $\mathbb{A}$ passing through it is in orange; events $a_{\pm n}$ and $a_{\pm m}$ are blue and pink ($n=160$, $m=40$); the sets of $b^{\pm}$ events that are $m$ links away to the future and past from $a_0$ are in brown, if they are at proper spatial distances larger than $0.2$ from $a_{\pm m}$, and in blonde, if they are closer than $0.2$ to $a_{\pm m}$; the brown curves are the curves at constant proper time $\alpha_2t_Pm$ from $a_0$ in $\mathbb{M}^2$; and the blue curves are the curves at constant proper spatial distance $0.2$ from $a_{\pm m}$ in $\mathbb{M}^2$. The thick orange events form the set $\mathbb{A}_n$, while the blonde events form the sets $L_m^{\pm}$ with $N_m^\pm=100$.
}
\label{fig:Alembertian}
\end{figure}

A geodesic time-like curve in a causal set is a numerable sequence of ordered events
\begin{equation}
\mathbb{A}=\{a_i\in \mathcal{C} | \cdots \prec a_{-1}\prec a_0\prec a_{+1}\prec \cdots \} 
\end{equation}
such that for any pair of events $a_j \prec a_k \in \mathbb{A}$, the sequence $\{a_j,a_{j+1},\cdots,a_k\}$ is a maximal chain connecting the two events. Let us now define a geodesic segment $\mathbb{A}_n$ to be a finite subset of events in $\mathbb{A}$,
\begin{equation}
\mathbb{A}_n=\{a_{-n} \prec \cdots \prec a_{-1}\prec a_0\prec a_{1}\prec \cdots \prec a_{n} \}, 
\end{equation}
consisting of $2n+1$ events, $n$ of which are in the past of $a_0$, and the other $n$ are in its future, so $n_{\mathcal{C}}(a_0,a_n)=n_{\mathcal{C}}(a_{-n},a_0)=n$. An example of such a segment $\mathbb{A}_n$ is shown in orange in Fig.~\ref{fig:Alembertian}, where Mr.~White~$a_0$ is at the origin. The boundary elements $a_{\pm n}$ of $\mathbb{A}_n$ are Mr.~Blue.

Equations~(\ref{eq:tauc},\ref{error_tau}) say that the proper time from $a_0$ to $a_{\pm n}$ is $\tau_{{\mathbb M}^{d+1}}(a_0,a_{\pm n})\approx \alpha_d t_P n$, with an error term $\sim t_P^{1-\beta_d}$ that goes to zero in the limit $t_P \rightarrow 0$. Using this, we show in Appendix~\ref{app:geo} that there exists a reference frame $R_{\mathbb{A}}$ in $\mathbb{M}^{d+1}$ in which the event $a_0$ is at the origin, while the temporal and spatial radial coordinates of $a_{\pm n}$ are 
\begin{align}\label{eq:tan}
t(a_{\pm n})&=\pm \alpha_d t_P n+\mathcal{O}\left(t_P^{1-\beta_d}\right),\\
r(a_{\pm n}) &\sim \sqrt{t_P^{2-\beta_d} n}.
\label{eq:ran}
\end{align}

We now want our segment $\mathbb{A}_n$ to be \emph{local}. Our definition of what it means for it to be \emph{local} is that the temporal and spatial coordinates of its boundaries $a_{\pm n}$ must go to zero in the continuum limit. In view of the last two equations, this requirement is
\begin{align}
  t_Pn&\ll 1,\label{eq:temploc}\\
  t_P^{2-\beta_d}n&\ll 1.
\end{align} 
However, since $\beta_d<1$, the former requirement implies the latter. 

\subsubsection{Spatial neighborhoods}\label{ssec:neighbor-spatial}

The basic idea behind our definition of local neighborhoods is   locality not only in time, but also in space. That is, we want to define a set of events that are at approximately the same small proper time from $a_0$, but also lie within a small spatial distance from one of such events.

Let this event be $a_m\in\mathbb{A}_n$, $m \le n$, Mr.~Pink in Fig.~\ref{fig:Alembertian}. This is one of the infinite number of brown events that are $m$ links to the future of $a_0$ in $\cal{C}$. Let us denote them all by $b^+$. By construction, they are space-like separated among themselves, and are at proper time
\begin{equation}
  \tau_m = \alpha_d t_P m + \O\left( t_P^{1-\beta_d}\right)
\end{equation}
from $a_0$, which follows from Eqs.~(\ref{eq:tauc},\ref{error_tau}). Henceforth, to make sure that the error term in the last equation is indeed negligible, we require that
\begin{equation}\label{eq:m-lower-scaling}
m\gg t_P^{-\beta_d}\text{ or }\tau_m\gg t_P^{1-\beta_d}.
\end{equation}

Let us now sort the $b^+$ events in the order of increasing proper spatial distance from the reference point~$a_m$, and define the set $L^+_m$, Mr.~Blonde in Fig.~\ref{fig:Alembertian}, to be the set of $N^+_m$ events $b^+$ that are spatially closest to $a_m$:
\begin{equation}
L^+_m \equiv \{ b^{\scriptscriptstyle+}_i \in \mathcal{C} \, | \, d(b_1^{\scriptscriptstyle+},a_{m})< d(b_2^{\scriptscriptstyle+},a_{m})< \cdots <d(b_{N^{\scriptscriptstyle+}_m}^{\scriptscriptstyle+},a_{m})\}.
\end{equation}
This definition makes sense because it is possible to correctly measure proper lengths at all scales based only on the causal set structure~\cite{Boguna:2024qf}.

In Appendix~\ref{app:N+} we calculate $N^+_m$ as a function of the radius $l_c$ of $L^+_m$, defined as the proper distance from $a_m$ to the boundary event in $L^+_m$. The sets $b^-$ and $L^-_m$ are defined similarly, except that they are to the past of~$a_0$.

\subsubsection{Local neighborhoods}

Finally, we define the \emph{local neighborhood} of event $a_0$ to be the union of these three sets: $\mathbb{A}_n$, $L_m^+$ and $L_m^-$, with some proper scaling of the sizes $N_m^\pm$ of $L_m^\pm$ that makes them \emph{local}, which we determine in Section~\ref{ssec:terms}.

We stress that this definition is Lorentz invariant since it is based only on Lorentz invariant quantities---proper times and lengths---and that these quantities are measured using only the causal set structure.\\

\subsection{Measuring temporal derivatives $\partial_t^2 \phi$}\label{ssec:temporal}

In this section we introduce the following discrete second-order time derivative operator in causal sets:
\begin{equation}\label{eq:Dt2}
  \D_t \phi(a_0) = \frac{\phi(a_n)+\phi(a_{-n})-2 \phi(a_0)}{ (\alpha_d t_P n)^2},
\end{equation}
and show that it converges to the continuous second-order time derivative in the continuum limit:
\begin{equation}
  \lim_{t_P\to0} \D_t \phi(a_0) = \left.\frac{\partial^2 \phi}{\partial t^2}\right|_{a_0}.
\end{equation}

To show this, we first evaluate the scalar field $\phi$ at the boundary events $a_{\pm n}$ of the local temporal segment $\mathbb{A}_n$ from Section~\ref{ssec:neighbor-temporal} via the Taylor series expansion around the origin event $a_0$ in the reference frame $R_{\mathbb{A}}$:
\begin{widetext}
\begin{equation}
\phi(a_{\pm n}) \approx \phi(a_0)+\left.\frac{\partial \phi}{\partial t}\right|_{a_0} t(a_{\pm n})+\left.\nabla{\phi}\right|_{a_0} \cdot \vec{r}(a_{\pm n})+\frac{1}{2} \left.\frac{\partial^2 \phi}{\partial t^2}\right|_{a_0} t^2(a_{\pm n}),
\label{eq:phia}
\end{equation}
\end{widetext}
where $t(a_{\pm n})$ and $\vec{r}(a_{\pm n})$ are the temporal and spatial coordinates of $a_{\pm n}$ in $R_{\mathbb{A}}$.
Using this expansion, we now evaluate the sum $\phi(a_n)+\phi(a_{-n})$ appearing in Eq.~\eqref{eq:Dt2}. This sum involves the first-order time derivative term with factor $\left[ t(a_{n})+t(a_{-n})\right]$, which is $\sim t_P^{1-\beta_d}$ according to Eq.~\eqref{eq:tan}. The spatial gradient term in the sum comes with factor $\sim \sqrt{t_P^{2-\beta_d} n}$ according to Eq.~\eqref{eq:ran}. Finally, the factor $\left[ t^2(a_{n})+t^2(a_{-n})\right]/2$ of the second-order time derivative term in the sum is $\approx (\alpha_d t_P n)^2$ according to Eq.~\eqref{eq:tan}. Among these three factors---$\,\sim t_P^{1-\beta_d}$, $\sim \sqrt{t_P^{2-\beta_d} n}$, and $\sim t_P^2 n^2$---the last one is dominating if $n\gg t_P^{-\frac{2+\beta_d}{3}}$. One can also check that higher-order derivative terms are negligible. Among those, the largest is the fourth-order time derivative. It is $t_P^4 n^4$, which is $\ll t_P^2n^2$ according to Eq.~\eqref{eq:temploc}. Combining these considerations, we conclude that
\begin{equation}
\left.\frac{\partial^2 \phi}{\partial t^2}\right|_{a_0}=\D_t\phi(a_0)+\O\left(t_P^{-\frac{2+\beta_d}{2}} n^{-\frac{3}{2}}\right)+\O\left(t_P^2 n^2 \right),
\label{eq:partial2t}
\end{equation}
where the first and second error terms come, respectively, from the gradient and fourth-order time derivative terms in the Taylor series expansion of $\phi(a_0)$. Both error terms go to zero in the continuum limit $t_P\to0$ if
\begin{equation}
t_P^{-\frac{2+\beta_d}{3}} \ll n \ll t_P^{-1},
\label{eq:range3}
\end{equation}
which is always possible since $\beta_d<1$, cf.~Section~\ref{ssec:proper-time}. There is also the optimal scaling of $n$ as a function of $t_P$ that minimizes the total error in Eq.~\eqref{eq:partial2t}. This scaling is
\begin{equation}
n \sim t_P^{-\frac{6+\beta_d}{7}},
\end{equation}
in which case Eq.~\eqref{eq:partial2t} becomes
\begin{equation}\label{eq:partial2t-optimal}
  \left.\frac{\partial^2 \phi}{\partial t^2}\right|_{a_0}=\D_t\phi(a_0)+\O\left( t_P^{\frac{2}{7}(1-\beta_d)} \right).
\end{equation}

\subsection{Measuring spatial derivatives $\nabla^2 \phi$}\label{ssec:spatial}

The last part of our discrete d'Alembertian construction is the definition of the discrete second-order spatial derivative in causal sets. The key idea behind this definition is similar to the one we used to construct the discrete Riemannian Laplacian in Section~\ref{sec:laplacian}. Specifically, we simply average the field $\phi$ over all the events in the spatial neighborhoods $L_m^\pm$ from Section~\ref{ssec:neighbor-spatial},
\begin{equation}
\overline{\phi^\pm}=\frac{1}{N^\pm_m} \sum_{i=1}^{N^\pm_m} \phi(b_i^\pm),
\label{eq:phiaverage}
\end{equation}
and then relate the difference
\begin{equation}\label{eq:spatial-protoboxop}
\overline{\phi^+}+\overline{\phi^-}-2\phi(a_0)
\end{equation}
to~$\nabla^2 \phi$.

To do so, we consider the field value $\phi(b^+_i)$ at event $b^+_i \in L^+_m$ as the Taylor series expansion of the field at $a_0$:
\begin{widetext}
\begin{equation}
\phi(b^+_i) \approx \phi(a_0)+\left.\frac{\partial \phi}{\partial t}\right|_{a_0} t^+_i+\left.\nabla{\phi}\right|_{a_0} \cdot \vec{r}^+_i+\frac{1}{2} 
\left.\partial_{\mu \nu}^2 \phi \right|_{a_0} x_i^{+\mu}x_i^{+\nu} + \O\left(x_i^{+\mu}x_i^{+\nu}x_i^{+\lambda}\right),
\label{eq:phiplus}
\end{equation}
\end{widetext}
and similarly for $b^-_i \in L^-_m$, where $t_i^+$ and $\vec{r}^+_i$ are the temporal and spatial coordinates of $b_i^+$ in $R_{\mathbb{A}}$, and $x_i^+=(t_i^+,\vec{r}^+_i)$. This expansion is possible if \begin{equation}\label{eq:local,really}t_i^\pm \ll 1\text{ and }r_i^\pm \ll1\end{equation} for all $i=1,\cdots,N^\pm_m$, which imposes some conditions on $m$ that we determine in Section~\ref{ssec:terms}.

Assuming these conditions are satisfied, let us see what happens when we start averaging the terms on the right-hand side of Eq.~\eqref{eq:phiplus} over all events $b_i^+\in L_m^+$ to obtain the average field $\overline{\phi^+}$ in Eq.~\eqref{eq:phiaverage}. The resulting averages are population averages over the set of events in $L_m^+$. However, since these events are random, their population averages are also random. They are random variables that depend on a particular realization of the Poisson point process. As in the previous sections, we denote averages over the sprinkling process by brackets~$\langle \cdot \rangle$, whereas $\overline{\text{overbars}}$ denote population averages.

\subsubsection{First-order derivatives}

We start by analyzing the population average of the term with the first-order temporal derivative $\partial_t\phi$ in Eq.~\eqref{eq:phiplus}. It comes with the factor
\begin{equation}
\overline{t^+} \equiv \frac{1}{N_m^+} \sum_{i=1}^{N_m^+}  t^+_i.
\label{eq:tplusaverage}
\end{equation}
The population average $\overline{t^+}$ is a random variable whose average $\langle \overline{t^+} \rangle$ over the Poisson point process is $\langle t^+ \rangle$, where $t^+$ is the temporal coordinate of an event chosen uniformly at random from the random set $L_m^+$, and $\langle t^+ \rangle$ is its average. Since events in $L_m^+$ are uncorrelated, the variance of $\overline{t^+}$ is given by
\begin{equation}
\text{Var}(\overline{t^+})=\frac{\text{Var}(t^+)}{N_m^+}.
\label{eq:vartplus}
\end{equation}
Combining these observations, we write that
\begin{equation}
\overline{t^+}=\langle t^+ \rangle+\O\left(\sqrt{\frac{\text{Var}(t^+)}{N_m^+}} \right).
\label{eq:tplus}
\end{equation}

Next, due to the rotational symmetry of $\mathbb{M}^{d+1}$, the population averages of terms with odd powers of any spatial coordinate are random variables with zero mean over the Poisson point process. These include, in the first place, the term with the field gradient $\nabla\phi$ in the right hand side of Eq.~\eqref{eq:phiplus}, which comes with the factor
\begin{equation}
\overline{\vec{r}^+} \equiv \frac{1}{N_m^+} \sum_{i=1}^{N_m^+}  \vec{r}^+_i.
\end{equation}
The average coordinate $\langle \vec{r}^+ \rangle$ of a random node in $L_m^+$ is $\vec{0}$, and therefore $\langle \overline{\vec{r}^+} \rangle=\vec{0}$. Since events in $L_m^+$ are uncorrelated, the variance of each individual component of $\overline{\vec{r}^+}$---let it be the $x$ component for concreteness---is given by
\begin{equation}\label{eq:gradientvariance}
\text{Var}(\overline{x^+})=\frac{\langle x^{+2}\rangle}{N_m^+},
\end{equation}
where $x^+$ is the $x$-coordinate of a random node in $L_m^+$, and where we have used that $\langle x^+ \rangle=0$. Invoking the symmetry of $\mathbb{M}^{d+1}$ again, we also note that the second moment of any spatial coordinate is the same and equal to $\langle x^{+2}\rangle$. Therefore, all the components of vector $\overline{\vec{r}^+}$ are of the order 
\begin{equation}\label{eq:gradientaverage}
\overline{x^+} \sim \O\left(\sqrt{\frac{\langle x^{+2}\rangle}{N_m^+}}\right).
\end{equation}

\subsubsection{Second-order derivatives}

We now focus on the terms with the second-order derivatives $\partial_{\mu \nu}^2 \phi$ of the field in Eq.~\eqref{eq:phiplus}. Consider first the second-order temporal derivative $\partial_t^2\phi$. This term comes with the factor
\begin{equation}
\overline{t^{+2}} \equiv \frac{1}{N_m^+} \sum_{i=1}^{N_m^+}t^{+2}_i.
\label{eq:tplusaverage2}
\end{equation}
This is a random variable whose average $\langle \overline{t^{+2}} \rangle$ over the Poisson point process is $\langle t^{+2} \rangle$. Therefore, this term dominates the gradient term with factors in Eq.~\eqref{eq:gradientaverage} if $\langle t^{+2} \rangle \gg \sqrt{\langle x^{+2}\rangle/N_m^+}$, or equivalently, 
\begin{equation}
N_m^+ \gg \frac{\langle x^{+2}\rangle}{\langle t^{+2}\rangle^2}.
\label{eq:Ncondition1}
\end{equation}
This is an important condition on the number of events in $L_m^+$: if it holds, the second-order derivative term dominates the gradient term in Eq.~\eqref{eq:phiplus}, so the latter becomes a subleading error term. The evaluation of this condition calls for the evaluation of $\langle t^{+2}\rangle$ and $\langle x^{+2}\rangle$, which we do in Appendix~\ref{app:t+}, a gist of which appears in Section~\ref{ssec:terms}, where we collect all other error terms. We also note at this point, that we do not have to worry about the domination of the second-order derivative terms over the first-order time derivative one, since the latter, coming with factor $\sim\langle t^{+}\rangle$, will be canceled in Eq.~\eqref{eq:spatial-protoboxop} by its past counterpart from $L_m^-$, coming with factor $\sim\langle t^{-}\rangle = -\langle t^{+}\rangle$, as we further discuss in Section~\ref{ssec:terms} as well.

We now move to the second-order spatial derivatives in Eq.~\eqref{eq:phiplus}. Let us consider the $\partial_x^2\phi$ term. It comes with the factor
\begin{equation}
\overline{x^{+2}} \equiv \frac{1}{N_m^+} \sum_{i=1}^{N_m^+}x^{+2}_i,
\end{equation}
whose sprinkling average $\langle \overline{x^{+2}} \rangle$ is $\langle x^{+2} \rangle$. Therefore, according to Eq.~\eqref{eq:gradientaverage}, this term dominates the gradient term in Eq.~\eqref{eq:phiplus} if $\langle x^{+2} \rangle \gg \sqrt{\langle x^{+2}\rangle/N_m^+}$, or equivalently, 
\begin{equation}
N_m^+ \gg \frac{1}{\langle x^{+2}\rangle},
\label{eq:Ncondition2}
\end{equation}
which is another condition that $N_m^+$ must satisfy. We note that according to Appendix~\ref{app:t+}, the values of $\langle x^{+2}\rangle$ and $\langle t^{+2}\rangle$ are of the same order of magnitude, so that both the second-order temporal and spatial derivative terms in Eq.~\eqref{eq:phiplus} are equally important.

Finally, the second-order derivative terms with mixed derivatives like $\partial_t\partial_x\phi$ or $\partial_x\partial_y\phi$ come with factors like
\begin{align}
\overline{t^{+}x^+} &\equiv \frac{1}{N_m^+} \sum_{i=1}^{N_m^+}t^{+}_i x^+_i,\\
\overline{x^{+}y^+} &\equiv \frac{1}{N_m^+} \sum_{i=1}^{N_m^+}x^{+}_i y^+_i
\end{align}
and similarly for the rest of the mixed terms. Once again the symmetry of $\mathbb{M}^{d+1}$ implies that $\langle \overline{t^{+}x^+} \rangle=0$ and $\langle \overline{x^{+}y^+} \rangle=0$, and their variances 
\begin{align}
\text{Var}(\overline{t^{+}x^+})&=\frac{\langle t^{+2} x^{+2}\rangle}{N_m^+},\\
\text{Var}(\overline{x^{+}y^+})&=\frac{\langle x^{+2} y^{+2}\rangle}{N_m^+}
\end{align}
are negligible compared to the variance Eq.~\eqref{eq:gradientvariance} of the gradient term because $\langle t^{+2} x^{+2}\rangle \ll \langle x^{+2} \rangle$ and $\langle x^{+2} y^{+2}\rangle \ll \langle x^{+2} \rangle$ if $t_i^+$, $x_i^+$, and $y_i^+$ are $\ll1$ for all events $b_i^+$ in $L_m^+$ ($i=1,\cdots,N_m^+$). That is, all these terms are negligible with respect to the gradient term in Eq.~\eqref{eq:phiplus}.

\subsubsection{Higher-order derivatives}

The higher-order derivative terms in  Eq.~\eqref{eq:phiplus} come with factors that are population averages of higher powers of temporal and spatial coordinates, or only spatial coordinates, of events in $L_m^+$. Let us analyze one class of such terms as an example. Let their factors be
\begin{equation}
\overline{x^{+l}t^{+k}}\equiv \frac{1}{N_m^+} \sum_{i=1}^{N_m^+}x^{+l}_i t^{+k}_i \; \; \text{with} \; \; l+k\ge 3.
\end{equation}
The symmetry of $\mathbb{M}^{d+1}$ implies that the ensemble average of terms with odd $l$ values is zero. Similarly, due to the symmetry between the sets $L_m^+$ and $L_m^-$, all terms with odd $k$ values have zero ensemble average once averaged over the two sets. This implies that the contribution of these terms to the sum $\overline{\phi^+}+\overline{\phi^-}$ reduces to their statistical error, through their variance
\begin{equation}
\text{Var}(\overline{x^{+l}t^{+k}})=\frac{\text{Var}( x^{+l} t^{+k})}{N_m^+},
\end{equation}
which is negligible compared to the variance Eq.~\eqref{eq:gradientvariance} of the gradient term because $\text{Var}( x^{+l} t^{+k})<\langle x^+ \rangle$.

The only higher-order terms with non-vanishing ensemble average are those with even values of both $l$ and $k$. However, thanks to our locality requirement $t_i^+ \ll 1$ and $x_i^+ \ll1$ for all $i=1,\cdots,N_m^+$, these terms are subdominant with respect to the second-order derivative terms.
Therefore, out of these higher-order terms, we consider only the largest fourth-order terms in our recollection of all the terms in the next section. 

\subsubsection{Collecting all the terms and scalings}\label{ssec:terms}

Here we first recollect all the scaling requirements to $m$, $\tau_m$, and $N_m^\pm$. The first ones are the locality requirements $t_i^{+} \ll 1$ and $x_i^{+} \ll 1$ in Eq.~\eqref{eq:local,really} for the temporal and spatial coordinates of all events $b_i^+$ in the local spatial neighborhood $L_m^+$ ($i=1,\cdots,N_m^+$). We show in Appendix~\ref{app:N+} that these requirements impose the following constrains on $\tau_m$ and $L_m^+$'s radius $l_c$ (the proper length from $a_m$ to the farthest event in $L_m^+$):
\begin{equation}
  \tau_m \ll 1\text{ and }l_c \ll \sqrt{\tau_m}.
\end{equation}

We also have the basic asymptotic requirement
\begin{equation}\label{eq:taum-vs-tp}
\tau_m\gg t_P^{1-\beta_d}
\end{equation}
from Eq.~\eqref{eq:m-lower-scaling}, which we use in Appendix~\ref{app:N+} to obtain an approximate expression for $N_m^+$ as a function of $\tau_m$ and $l_c$. Since we have some freedom in choosing how $l_c$ scales with respect to $\tau_m$, we show in Appendix~\ref{app:N+} that we can use this freedom to let $l_c$ scale proportionally to $\tau_m$,
\begin{equation}\label{eq:lc-vs-taum}
l_c\sim\tau_m,
\end{equation}
without violating any scaling requirements. This setting simplifies all the expressions significantly. In particular, the expression for $N_m^+$ simplifies to
\begin{equation}\label{eq:nm-simple}
  N^+_m\sim t_P^{-(d+\beta_d)} \tau_m^d.
\end{equation}

Yet we also have the critical requirements in Eqs.~(\ref{eq:Ncondition1},\ref{eq:Ncondition2}). They say that the number $N_m^+$ of events in the local neighborhood $L_m^+$ must be sufficiently high to ensure that the ensemble averages of terms involving first-order derivatives are negligible compared to those involving second-order derivatives. These conditions involve $\langle t^{+2}\rangle$ and $\langle x^{+2}\rangle$. We calculate these quantities in Appendix~\ref{app:t+}, and show that under the conditions in Eqs.~(\ref{eq:taum-vs-tp},\ref{eq:lc-vs-taum}), they simplify to
\begin{equation}\label{eq:t2-x2-taum2}
\langle t^{+2} \rangle \sim \langle x^{+2} \rangle \sim \tau_m^2.
\end{equation}
Therefore, the requirements in Eqs.~(\ref{eq:Ncondition1},\ref{eq:Ncondition2}) become simply $N_m^+\gg1/\tau_m^2$, which in view of Eq.~\eqref{eq:nm-simple} is equivalent to
\begin{equation}
  \tau_m \gg t_P^\frac{d+\beta_d}{d+2}.
\end{equation}
This requirement is stronger than the one in Eq.~\eqref{eq:taum-vs-tp}, given the values of $\beta_d$ summarized in Section~\ref{ssec:proper-time}. Therefore, under the assumption in Eq.~\eqref{eq:lc-vs-taum}, the final allowed scalings of $\tau_m$ and $m$ satisfying all the requirements above are
 \begin{align}
\label{eq:range0}
t_P^{\frac{d+\beta_d}{d+2}} &\ll \tau_m \ll 1,\\
t_P^{-\frac{2-\beta_d}{d+2}} &\ll m \ll t_P^{-1},
\label{eq:range1}
\end{align} 
which can always be satisfied in the continuum limit $t_P \rightarrow 0$.

We now assume that the scaling relations in Eqs.~(\ref{eq:lc-vs-taum}-\ref{eq:range1}) are enforced, and under this assumption, we summarize in Table~\ref{table:allorders} all the results in this Section~\ref{ssec:spatial}. The table recollects all the terms, up to the fourth order, appearing in Eq.~\eqref{eq:phiplus}, along with their population-averaged and ensemble-averaged factors and their error corrections emerging from Eq.~\eqref{eq:phiaverage}. In the table, we use the fact that, as a consequence of the results in Appendix~\ref{app:t+}, $\langle x^{+l}t^{+k} \rangle \sim \tau_m^{l+k}$ if $l$ is even, and $\langle x^{+l}t^{+k} \rangle = 0$ if $l$ is odd due to the rotational symmetry. 

\begin{table*}
  \centering
  \caption{The terms in the Taylor series expansion of the field in Eq.~\eqref{eq:phiplus} up to the fourth order, and the expectations and error corrections of their factors emerging from the averaging in Eq.~\eqref{eq:phiaverage}. The shown values assume the scaling relations in Eqs.~(\ref{eq:lc-vs-taum}-\ref{eq:range1}). The terms highlighted in the blue color are the ones contributing to the final expression in Eq.~\eqref{eq:nabla}. The third- and fourth-order terms with zero ensemble averages are omitted.}
  \label{table:allorders}
  \begin{tabular}{|c|c|c|c|} 
    \hline \hline
    Term & Factor's population average & Ensemble average & Error correction \\\hline
    $t_i\partial_t\phi$ & $\overline{t^+}$ & ${\langle t^+ \rangle \sim \tau_m}$  & $\blue{\O\left(\frac{\tau_m}{\sqrt{N_m^+}}\right)}$  \\[0.3cm]
    $x_i\partial_x\phi,y_i\partial_y\phi,z_i\partial_z\phi$ & $\overline{x^+},\overline{y^+},\overline{z^+}$  & $0$  & $\blue{\O\left(\frac{\tau_m}{\sqrt{N_m^+}}\right)}$ \\[0.3cm] 
    $t_i^2\partial_t^2\phi$ & $\overline{t^{+2}}$ & $\blue{\langle t^{+2} \rangle \sim \tau_m^2}$ & $\O\left(\frac{\tau_m^2}{\sqrt{N_m^+}}\right)$ \\[0.3cm]
    $t_ix_i\partial_t\partial_x\phi,t_iy_i\partial_t\partial_y\phi,t_iz_i\partial_t\partial_z\phi$ & $\overline{t^{+}x^+},\overline{t^{+}y^+},\overline{t^{+}z^+}$ & $0$ & $\O\left(\frac{\tau_m^2}{\sqrt{N_m^+}}\right)$ \\[0.3cm]
    $x_iy_i\partial_x\partial_y\phi,y_iz_i\partial_y\partial_z\phi,z_ix_i\partial_z\partial_x\phi$ & $\overline{x^{+}y^+},\overline{y^{+}z^+},\overline{z^{+}x^+}$ & $0$ & $\O\left(\frac{\tau_m^2}{\sqrt{N_m^+}}\right)$ \\[0.3cm]
    $x_i^2\partial_x^2\phi,y_i^2\partial_y^2\phi,z_i^2\partial_z^2\phi$ & $\overline{x^{+2}},\overline{y^{+2}},\overline{z^{+2}}$ & $\blue{\langle x^{+2} \rangle \sim \tau_m^2}$ & $\O\left(\frac{\tau_m^2}{\sqrt{N_m^+}}\right)$ \\[0.3cm]
    $t_i^3\partial_t^3\phi$ & $\overline{t^{+3}}$ & $\langle t^{+3} \rangle \sim \tau_m^3$ & $\O\left(\frac{\tau_m^3}{\sqrt{N_m^+}}\right)$ \\[0.3cm]
    $t_ix_i^2\partial_t\partial_x^2\phi,t_iy_i^2\partial_t\partial_y^2\phi,t_iz_i^2\partial_t\partial_z^2\phi$ & $\overline{t^+x^{+2}},\overline{t^+y^{+2}},\overline{t^+z^{+2}}$ & $\langle t^+x^{+2} \rangle \sim \tau_m^3$ & $\O\left(\frac{\tau_m^3}{\sqrt{N_m^+}}\right)$ \\[0.3cm]
    $t_i^4\partial_t^4\phi$ & $\overline{t^{+4}}$ & $\color{blue}{\langle t^{+4} \rangle\sim\tau_m^4}$ & $\O\left(\frac{\tau_m^4}{\sqrt{N_m^+}}\right)$ \\[0.3cm]
    $t_i^2x_i^2\partial_t^2\partial_x^2\phi,t_i^2y_i^2\partial_t^2\partial_y^2\phi,t_i^2z_i^2\partial_t^2\partial_z^2\phi$ & $\overline{t^{+2}x^{+2}},\overline{t^{+2}y^{+2}},\overline{t^{+2}z^{+2}}$ & $\color{blue}{\langle t^{+2}x^{+2} \rangle\sim\tau_m^4}$ & $\O\left(\frac{\tau_m^4}{\sqrt{N_m^+}}\right)$ \\[0.3cm]
    $x_i^2y_i^2\partial_x^2\partial_y^2\phi,y_i^2z_i^2\partial_y^2\partial_z^2\phi,z_i^2x_i^2\partial_z^2\partial_x^2\phi$ & $\overline{x^{+2}y^{+2}},\overline{y^{+2}z^{+2}},\overline{z^{+2}x^{+2}}$ & $\color{blue}{\langle x^{+2}y^{+2} \rangle\sim\tau_m^4}$ & $\O\left(\frac{\tau_m^4}{\sqrt{N_m^+}}\right)$ \\[0.3cm]
    $x_i^4\partial_x^4\phi,y_i^4\partial_y^4\phi,z_i^4\partial_z^4\phi$ & $\overline{x^{+4}},\overline{y^{+4}},\overline{z^{+4}}$ & $\color{blue}{\langle x^{+4} \rangle\sim\tau_m^4}$ & $\O\left(\frac{\tau_m^4}{\sqrt{N_m^+}}\right)$ \\[0.3cm]
    % ---------------------------------------
    \hline \hline
  \end{tabular}
\end{table*}

The leading terms in Table~\ref{table:allorders} are marked in blue. To see what terms are leading, we observe that in the scaling regime of Eqs.~(\ref{eq:lc-vs-taum}-\ref{eq:range1}),
\begin{align}
\tau_m &\gg \tau_m^2 \gg \frac{\tau_m}{\sqrt{N_m^+}},\\
\tau_m^2 &\gg \tau_m^4 \gg  \frac{\tau_m^2}{\sqrt{N_m^+}},
\end{align}
yet which one of the two terms,
\begin{equation}
\tau_m^4 \text{ versus }\frac{\tau_m}{\sqrt{N_m^+}},
\end{equation}
is leading is not known a priori since this depends on a particular scaling of $\tau_m$ as a function of $t_P$.

The fact that the terms with odd powers of $t^+$ in Table~\ref{table:allorders} are not in blue is because the table reports only the averages $\overline{\phi^+}$ over events $b_i^+\in L_m^+$, but we have to add $\overline{\phi^+}$ to $\overline{\phi^-}$ according to Eq.~\eqref{eq:spatial-protoboxop}. Upon this addition, due to the symmetry between $L_m^+$ and $L_m^-$ (we set $N^-_m=N^+_m$), the expectations with odd powers of $t^+$ cancel out, e.g., $\langle t^+ \rangle+\langle t^- \rangle=0$. Therefore, the contributions from these terms to $\overline{\phi^+}+\overline{\phi^-}$ reduce to their error corrections, which are subleading.

Recollecting all these observations and leading terms in Table~\ref{table:allorders}, we finally conclude that
\begin{align}
\label{eq:nabla}
\langle t^{+2}\rangle\left.\frac{\partial^2 \phi}{\partial t^2}\right|_{a_0}&+\langle x^{+2}\rangle\left.\nabla^2{\phi}\right|_{a_0}=
\overline{\phi^+}+\overline{\phi^-}-2\phi(a_0)\\&+\O\left(\frac{\tau_m}{\sqrt{N_m^+}}\right)+\O\left(\tau_m^4 \right),\nonumber
\end{align}
As a reminder, this equation assumes that all the scaling relations in Eqs.~(\ref{eq:lc-vs-taum}-\ref{eq:range1}) are satisfied.
As evident from Table~\ref{table:allorders}, the first error term in Eq.~\eqref{eq:nabla} arises from terms involving first-order derivatives, whereas the second error term comes from fourth-order derivatives. In general, the dominant error is the larger of these two terms. However, as is also the case with Eq.~\eqref{eq:partial2t}, there is an optimal scaling of $m$ for which both error terms are of the same order of magnitude. This scaling is \begin{equation}
m \sim t_P^{-\frac{6-\beta_d}{d+6}},
\label{eq:moptimal}
\end{equation} which lies in the regime of Eq.~\eqref{eq:range1}. With this scaling of~$m$, Eq.~\eqref{eq:nabla} becomes
\begin{align}\label{eq:nabla-optimal}
\langle t^{+2}\rangle\left.\frac{\partial^2 \phi}{\partial t^2}\right|_{a_0}&+\langle x^{+2}\rangle\left.\nabla^2{\phi}\right|_{a_0}=
\overline{\phi^+}+\overline{\phi^-}-2\phi(a_0)\\&+\O\left( t_P^{\frac{4(d+\beta_d)}{d+6}}\right),\nonumber
\end{align}
whereas its contribution to the error of the d'Alembertian becomes
\begin{align}\label{eq:nabla-optimal2}
\O\left( t_P^{\frac{2(d+\beta_d)}{d+6}}\right).
\end{align}

\subsection{Measuring the d'Alembertian}\label{ssec:box}

We finally have everything in place to define our discrete d'Alembertian. Combining Eqs.~\eqref{eq:nabla-optimal} and~(\ref{eq:partial2t-optimal},\ref{eq:Dt2}), we get
\begin{widetext}
\begin{equation}
\left.\Box \phi\right|_{a_0}=\left.\left(-\frac{\partial^2}{\partial t^2}+\nabla^2\right)\phi\right|_{a_0}=-\left(1+\frac{\langle t^{+2} \rangle}{\langle x^{+2} \rangle} \right) \frac{ \phi(a_n)+\phi(a_{-n})-2 \phi(a_0)}{ (\alpha_d t_P n)^2}+\frac{\overline{\phi^+}+\overline{\phi^-}-2 \phi(a_0)}{\langle x^{+2} \rangle}+\O\left(  t_P^{\frac{2}{7}(1-\beta_d)} \right),
\label{eq:Alembertian}
\end{equation}
\end{widetext}
where the error term is the maximum of the two optimal errors in Eqs.~\eqref{eq:partial2t-optimal} and~\eqref{eq:nabla-optimal2}.

Using the scaling relations in Eqs.~(\ref{eq:lc-vs-taum}-\ref{eq:range1}), we show in Appendix~\ref{app:t+} that $\langle x^{+2} \rangle$ can be set equal to $\tau_m^2/C$ for any constant $C>0$, in which case $\langle t^{+2} \rangle=(d/C+1)\tau_m^2$, so Eq.~\eqref{eq:Alembertian} becomes
\begin{equation}\label{eq:Alembertian-simple}
  \Box\phi(a_0) =  \B\phi(a_0) + \O\left(  t_P^{\frac{2}{7}(1-\beta_d)} \right),
\end{equation}
where our discrete d'Alembert operator $\B$ is defined as the following sum of the discrete second-order temporal and spatial derivatives: 
\begin{align}\label{eq:Alembertian-defined}
  \B &= -(C+d+1)\D_t + C\D_s,\text{ where}\\
  \D_t\phi(a_0)& = \frac{\phi(a_n)+\phi(a_{-n})-2 \phi(a_0)}{ (\alpha_d t_P n)^2},\\
  \D_s\phi(a_0)& = \frac{\overline{\phi^+}+\overline{\phi^-}-2 \phi(a_0)}{(\alpha_d t_P m)^2}.
\end{align}
Equation~\eqref{eq:Alembertian-simple} implies the desired convergence:
\begin{equation}\label{eq:Alembertian-convergence}
  \lim_{t_P\to0} \B\phi(a_0) = \Box\phi(a_0).
\end{equation}

It is important to note that the definition of the discrete d'Alembertian in Eq.~\eqref{eq:Alembertian-defined} is not unique. Here, we have chosen a symmetric version that uses information from both the future and past of event $a_0$. Another option is to define a retarded version that uses information only from the past of~$a_0$. We could also define the discrete temporal derivative~$\D_t$ by averaging over all the events in~$\mathbb{A}_n$, similar to the averaging over all the events in $L_m^\pm$ in the definition of the discrete spatial derivative~$\D_s$. We also note that the averaging in the definition of~$\D_s$ appears unavoidable as it is needed to suppress the contributions from the first-order derivative terms in Eq.~\eqref{eq:phiplus}.

\subsection{Numerical simulations}\label{ssec:sims}

Finally, we perform numerical simulations to confirm the convergence of our discrete d'Alembertian in Eq.~\eqref{eq:Alembertian-defined} to its continuous brother.

To do so, we apply the discrete d'Alembertian to causal sets in $\mathbb{M}^{2+1}$ generated at increasing densities ranging from $\rho = 10^4$ to $\rho = 3\times 10^6$. For each density $\rho$, we sprinkle events uniformly at random into a rectangular box in $\mathbb{M}^{2+1}$ with temporal coordinate $t \in [-1, +1]$ and spatial coordinates $x, y \in [-1/2, +1/2]$. The event $a_0$, at which the d'Alembertian of the field is evaluated, is at the origin. To select event $a_m$, we first need to fix $m$ for each value of $\rho$. To do so, we follow Eq.~\eqref{eq:moptimal}, and set $m \propto \rho^{23/96}$. The event $a_m$ is then $m$ links to the future of~$a_0$. There are many such events, but we select $a_m$ to be the one with the smallest radial coordinate to ensure that the boundaries of the local spatial neighborhood $L_m^+$ do not cross the boundaries of our simulation box. To define $L_m^+$, we set $l_c = 2 \tau_m/3$, where $\tau_m=m/\sqrt{2}\rho^{1/3}$. According to Eq.~\eqref{eq:xplusc} in Appendix~\ref{app:N+}, this choice bounds the radial coordinates of events in $L_m^+$ to $\sqrt{80}\tau_m/9$, again ensuring that all these events lie within our simulation box. The past spatial neighborhood $L_m^-$ is set up symmetrically, while for the temporal neighborhood, we set $n=m$, $a_n=a_m$, and $a_{-n}=a_{-m}$.
Finally, to evaluate the constant~$C$ in Eq.~\eqref{eq:Alembertian-defined}, we observe that according to Appendix~\ref{app:t+}, if $d=2$, then
\begin{align}
\langle t^{+2} \rangle &= \frac{1}{3}\left(\cosh^2\chi_c + \cosh\chi_c + 1\right)\tau_m^2,\\
\langle x^{+2} \rangle &= \frac{1}{6}\left(\cosh^2\chi_c + \cosh\chi_c - 2\right)\tau_m^2.
\end{align}
It follows then from Eq.~\eqref{eq:chic} that our choice of $l_c = 2 \tau_m/3$ sets $\cosh\chi_c = 11/9$, so that $\langle t^{+2} \rangle = 301\tau_m^2/243$ and $\langle x^{+2} \rangle = 29\tau_m^2/243$, that is $C=243/29$.

With these settings, we evaluate our discrete d'Alembertian in Eq.~\eqref{eq:Alembertian-defined} acting on the field
\begin{equation}
\phi(t,x,y) = e^{-a \tau^2} = e^{-a(t^2 - x^2 - y^2)},
\end{equation}
on which nonlocal d'Alembertians from Section~\ref{sec:nonlocal} do not converge. The continuous d'Alembertian applied to this field at the origin yields $\Box \phi(a_0)=6a$. We sample multiple values of~$a$ uniformly at random in $[0,1]$, and for each sampled value and each density $\rho$, we generate a different realization of the Poisson point process in the simulation box, and measure $\B \phi(a_0)$ in each such sprinkling.

\begin{figure}[t]
\centerline{\includegraphics[width=\columnwidth]{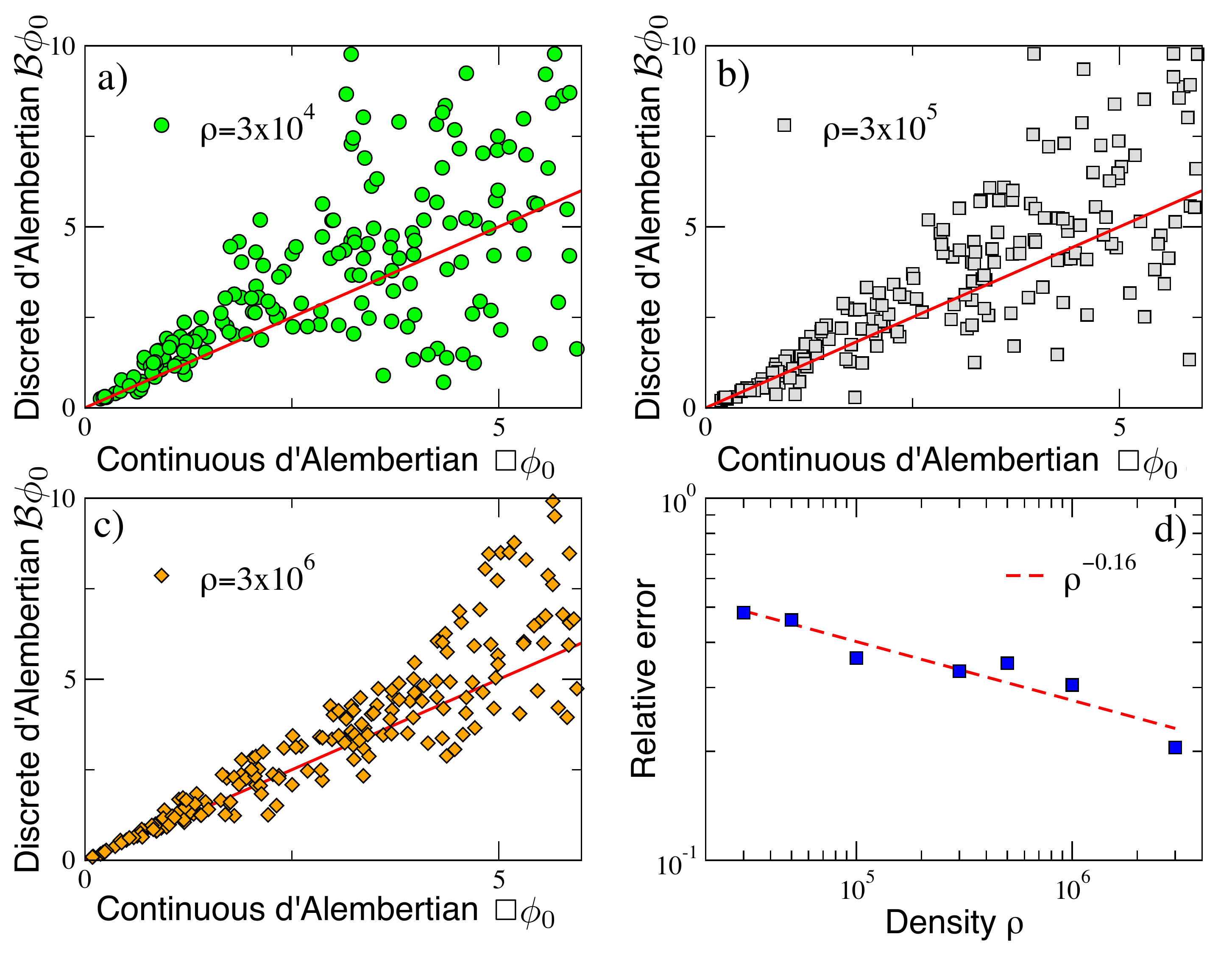}}
\caption{{\bf D'Alembertian simulations.} \textbf{(a-c)} The values of the discrete d'Alembertian $\B \phi(a_0)$ in Eq.~\eqref{eq:Alembertian-defined} acting on the field $\phi=e^{-a(t^2-x^2-y^2)}$ at $t=x=y=0$ in the $(2+1)$-dimensional Minkowski spacetime $\mathbb{M}^{2+1}$ vs.\ the continuous value $\Box \phi(a_0)=6a$ for $3$ different point densities~$\rho$ and $m=(2/5)\sqrt{2/a}\,\rho^{23/96}$. For each value of~$\rho$, $200$ random values of~$a\in[0,1]$ are used, and for each value of $a$, a separate realization of the Poisson point process is sampled. \textbf{(d)} The average relative error of the discrete vs.\ continuous d'Alembertian, defined as $\left\langle\left|\left(\B\phi(a_0)/\Box \phi(a_0)\right) - 1\right|\right\rangle$, for different values of $\rho$. For each value of~$\rho$, the relative error is averaged of $200$ random values of $a$ and sprinkling  realizations---one sprinkling for each~$a$.
The red dashed line is the fit $\rho^{-0.16}$, close to the analytical prediction $\O\left(\rho^{-2 (d+\beta_d)/[(d+6)(d+1)]}\right)=\O\left(\rho^{-3/16}\right)\approx\O\left(\rho^{-0.19}\right)$ from Eq.~\eqref{eq:nabla-optimal2}.
(Equation~\eqref{eq:nabla-optimal2} (vs.~\eqref{eq:Alembertian-simple}) yields the error scaling in this case because the field has the zero first-order derivatives at the origin, so Eq.~\eqref{eq:partial2t} contributes error only via its second error term, which is equivalent to the error term in Eq.~\eqref{eq:nabla-optimal2} since $n=m$.)
}
\label{fig9}
\end{figure}

Figure~\ref{fig9} shows the results. Panels~(a-c) juxtapose the measured values of the discrete d'Alembertian $\B \phi(a_0)$ against the continuous values $\Box \phi(a_0)=6a$ for $200$ different values of~$a$ and three different point densities~$\rho$. We observe that at higher densities, the discrete d'Alembertian values are concentrated more tightly around the continuous values.
Panel~(d) illustrates this further by showing the relative error between the discrete and continuous values. As predicted by our analysis in the previous sections, this error goes to zero polynomially as a function of the point density, implying the convergence.

\section{Concluding remarks}
\label{sec:conclusions}

Lorentzian geometry differs fundamentally from Riemannian geometry. In Riemannian geometry, the triangle inequality ensures that pairs of points that are close to a third point are also close to each other. In Lorentzian geometry, two points at a given proper time from a third point can be arbitrarily far apart from each other spatially. This observation suggests that causal set theory is inherently nonlocal, and that fundamental differential operators in the theory, including the d'Alembertian, should be nonlocal as well.

In this work, we demonstrated that causal set theory can, in fact, be formulated as a local theory, with discrete differential operators defined via a careful construction of local neighborhoods in a causal set. Based on this insight, we defined a local discrete d'Alembert operator, which converges the continuous \da in the continuum limit, which we have shown both analytically and numerically.

In addition to convergence, this operator has several other nice properties. First, since it operates on local neighborhoods, it converges on any field, as opposed to nonlocal operators that converge only on local fields that have compact supports. Second, our operator is self-averaging because when it acts on any field, the deviation of its values from their expectations in a single realization of the sprinkling process converges to zero in the continuum limit $\rho \rightarrow \infty$.
Finally, our operator is manifestly Lorentz-invariant because everything in its definition, including the local neighborhood, is formulated solely in terms of Lorentz-invariant quantities, such as proper times and proper lengths, which can be measured using only the causal set structure.

Generalizing, these results demonstrate that discrete approximations to continuous objects whose definitions are based on the notion of locality---by these we mean differential operators in the first place---are possible even in inherently nonlocal theories, of which causal set theory is a good example. It is thus plausible that similar ideas might help defining and proving convergence of discrete versions of more sophisticated locality-based objects, like spacetime curvature appearing in the Einstein field equations.

\begin{acknowledgments}
We thank Yasaman Yazdi for interesting discussions on a preliminary version of this work.
M.B.\ acknowledges support from grant TED2021-129791B-I00 funded by MCIN/AEI/10.13039/501100011033 and the ``European Union NextGenerationEU/PRTR'', Grant PID2022-137505NB-C22 funded by MCIN/AEI/10.13039/501100011033; Generalitat de Catalunya grant number 2021SGR00856. M.B.\ acknowledges the ICREA Academia award, funded by the Generalitat de Catalunya. D.K.\ acknowledges NSF Grant Nos.\ IIS-1741355 and CCF-2311160. 
\end{acknowledgments}

\appendix

\section{Numerical evaluation of constant $\alpha_2$}
\label{sec:alpha2}

\begin{figure}
\centerline{\includegraphics[width=0.8\columnwidth]{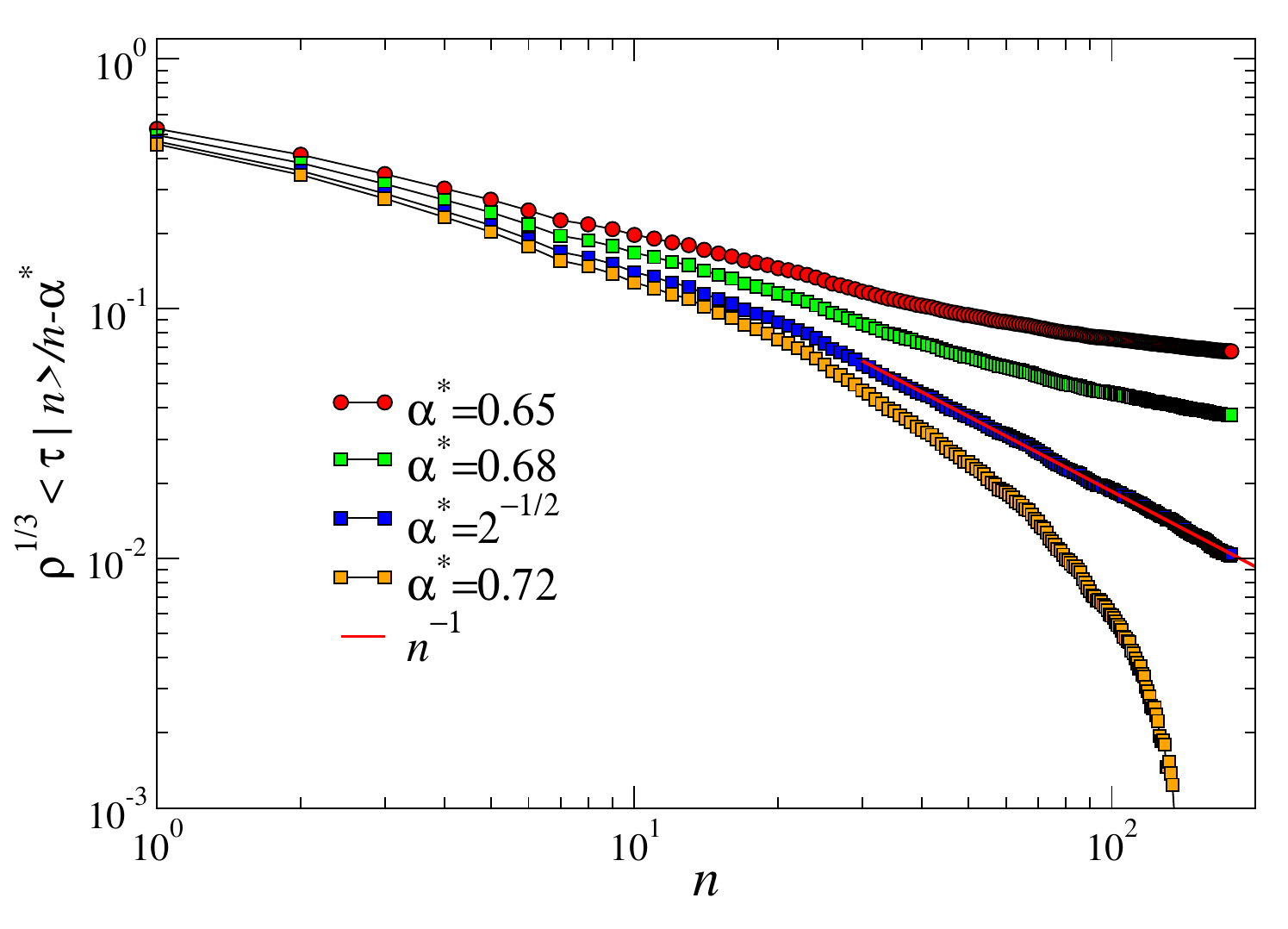}}
\caption{{\bf Estimation of parameter $\alpha_2$.} }
\label{fig:alpha2}
\end{figure}

Equation~\eqref{error_tau} allows us to estimate the value of the constant $\alpha_d$ in higher dimensions. By taking the average of $\tau_{{\mathbb M}^{d+1}}$ for a fixed number of causal set hops $n$ and an arbitrary parameter $\alpha^*$, we can write
\begin{equation}
\frac{\rho^{1/(d+1)}}{n}\langle \tau_{{\mathbb M}^{d+1}}| n \rangle-\alpha^*=\alpha_d-\alpha^*-\frac{\rho^{\frac{\beta_d}{d+1}} \langle \zeta_{d} \rangle}{n}.
\end{equation}
The left-hand side of this equation can be directly measured from numerical simulations. Therefore, the empirical value of $\alpha_d$ will be the value of $\alpha^*$ that makes the left-hand side behave as a perfect power law proportional to $n^{-1}$. Figure~\ref{fig:alpha2} shows results for causal sets generated by sprinkling events in ${\mathbb M}^{3}$ at density $\rho=3 \times 10^6$. As can be seen, the best power law is obtained when $\alpha^*=\alpha_2=1/\sqrt{2}$. Lower values of $\alpha^*$ lead to curves that bend upwards, whereas higher values result in curves decaying faster than a power law.

Unfortunately, in the case of $\mathbb{M}^4$, the maximum attainable value of $n$ grows as $\rho^{1/4}$. Therefore, to explore large values of $n$, we need to reach densities $\rho \sim 10^9$, which is beyond our current computational capability. However, preliminary results at a density of $\rho \sim 10^6$ support our conjecture that $\alpha_4=1/\sqrt{2}$.

\section{Fluctuations of geodesic paths in ${\mathcal{C}}$}
\label{app:geo}

Let $a$ and $b$ be two events in $\mathbb{M}^{d+1}$. We place event $a$ on the temporal axis at $(\tau_a,\vec{0})$, so its temporal coordinate is equal to the proper time from the origin to $a$. We place event $b$ with temporal coordinate $t_b < t_a$ and spatial radial offset $r_b$, such that the proper time from the origin to $b$ satisfies the equation $\tau_b^2 = t_b^2 - r_b^2$. If we now place a new reference frame with event $b$ at the origin of coordinates, the proper time between events $a$ and $b$ satisfies the equation $\tau_{ab}^2 = t_{ab}^2 - r_b^2$, with $t_{ab} = \tau_a - t_b$. Combining these three equations, we can express the spatial offset $r_b$ as a function of the three proper times as
\begin{widetext}
\begin{equation}
r_b^2 = \frac{1}{4\tau_a^2} \left[ (\tau_a + \tau_b + \tau_{ab})(\tau_a + \tau_b - \tau_{ab})(\tau_a - \tau_b + \tau_{ab})(\tau_a - \tau_b - \tau_{ab}) \right].
\label{eq:rb}
\end{equation}
\end{widetext}
On the other hand, Eq.~\eqref{error_tau} allows us to relate the proper time between two events in $\mathbb{M}^{d+1}$ and the proper time measured in the causal set by counting the number of hops in the geodesic path connecting both events plus an error term of order $t_P^{1-\beta_d}$. To simplify the notation, let us denote $\tilde{\tau} = \alpha_d t_P n$ as the proper time in $\mathcal{C}$, so the proper time in $\mathbb{M}^{d+1}$ can be expressed as $\tau = \tilde{\tau} + \delta \tau$, where $\delta \tau$ is the error term from~\eqref{error_tau}. The key point is to realize that if both events $a$ and $b$ belong to the same geodesic path in ${\mathcal{C}}$, then $\tilde{\tau}_a = \tilde{\tau}_b + \tilde{\tau}_{ab}$. Using this result in Eq.~\eqref{eq:rb}, we obtain
\begin{widetext}
\begin{equation}
r_b^2 = \frac{2\tilde{\tau}_a \tilde{\tau}_b \tilde{\tau}_{ab}}{(\tilde{\tau}_a + \delta \tau_a)^2}
\left( 1 + \frac{\delta \tau_a + \delta \tau_b + \delta \tau_{ab}}{2 \tilde{\tau}_a} \right)
\left( 1 + \frac{\delta \tau_a + \delta \tau_b - \delta \tau_{ab}}{2 \tilde{\tau}_b} \right)
\left( 1 + \frac{\delta \tau_a - \delta \tau_b + \delta \tau_{ab}}{2 \tilde{\tau}_{ab}} \right)
(\delta \tau_a - \delta \tau_b - \delta \tau_{ab}).
\label{eq:rb2}
\end{equation}
\end{widetext}
Notice that while $\delta \tau_b$ and $\delta \tau_{ab}$ are independent correction terms, the error $\delta \tau_a$ is not. Assuming that $\delta \tau / \tilde{\tau} \ll 1$ for $a$ and $b$, to first order in $\delta \tau$ we can write that
\begin{equation}
r_b^2 \approx 2\tilde{\tau}_b \left( 1 - \frac{\tilde{\tau}_b}{\tilde{\tau}_a} \right)(\delta \tau_a - \delta \tau_b - \delta \tau_{ab}).
\end{equation}
Since $\tilde{\tau}_b \le \tilde{\tau}_a$, the term within the parenthesis is bounded between zero and one, so for a general event $b$ different from $a$, its spatial offset is of order
\begin{equation}
r_b \sim \sqrt{\tilde{\tau}_b t_P^{1-\beta_d}}.
\end{equation}
Finally, using the relation between the proper and coordinate time of event $b$, we obtain that
\begin{equation}
t_b = \tilde{\tau}_b + O(t_P^{1-\beta_d}).
\end{equation}

 \section{Number of events in $L^\pm_m$}
 \label{app:N+}

Let us compute the expected number of events $N^+_m$ at a proper distance $l<l_c$ from the event $a_m \in L^+_m$. Suppose that $a_m$ is located at the position $(\tau_m,\vec{0})$ in $R_{\mathbb{A}}$. Thus, the region of $\mathbb{M}^{d+1}$ with proper length $l<l_c$ is bounded by the surface $r^2=l_c^2+(t-\tau_m)^2$, which in hyperbolic coordinates becomes
\begin{equation}
\cosh{\chi}=\frac{l_c^2+\tau^2+\tau_m^2}{2\tau \tau_m}.
\end{equation} 
The proper time of events in $L^+_m$ is also $\tau_m$. Therefore, such events have a hyperbolic coordinate $\chi<\chi_c$, where $\chi_c$ is the solution of
\begin{equation}
\cosh{\chi_c}=1+\frac{l_c^2}{2 \tau_m^2}.
\label{eq:chic}
\end{equation} 
Therefore, setting an upper bound for the proper length is equivalent to setting an upper bound for the coordinate~$\chi$, which translates to the following upper bounds for the temporal and spatial coordinates in the set $L_m^+$:
\begin{align}
t^+_c&=\tau_m \cosh{\chi_c}=\tau_m+\frac{l_c^2}{2 \tau_m},\\
x^+_c&=\tau_m \sinh{\chi_c}=l_c\sqrt{1+\frac{l_c^2}{4\tau_m^2}}.
\label{eq:xplusc}
\end{align}
From these equations, we see that the locality condition $t_i^+ \ll1$ and $x_i^+ \ll 1$ for $i=1,\cdots,N_m^+$ implies that
\begin{equation}\label{eq:taum-lc-scaling}
\tau_m \ll1 \; \text{and} \; l_c \ll \sqrt{\tau_m}.
\end{equation}
We note that we can choose any scaling of $l_c$ as long as these conditions are satisfied.

The volume element of $\mathbb{M}^{d+1}$ in hyperbolic coordinates is
\begin{equation}
dV=\tau^d d\tau \sinh^{d-1}{\chi}d \chi d \Omega_{d-1},
\end{equation}
where $d \Omega_{d-1}$ is the volume element of the $(d-1)$-sphere. The expected number of events from $L^+_m$ within an infinitesimal neighborhood of $(\tau, \chi, \Omega_{d-1})$ is then
\begin{equation}
dN^+=\rho \mbox{P}_d^+(\tau) \tau^d d\tau \sinh^{d-1}{\chi}d \chi d \Omega_{d-1},
\end{equation}
where $\mbox{P}_d^+(\tau)$ is the probability that an event at proper time $\tau$ from the origin belongs to the set $L^+_m$. The total number of expected events in $L^+_m$ is then given by
\begin{equation}
N^+_m=\rho \Omega_{d-1} \int_0^\infty \mbox{P}_d^+(\tau) \tau^d d\tau \int_0^{\chi_c}\sinh^{d-1}{\chi}d \chi.
\end{equation}
The probability $\mbox{P}_d^+(\tau)$ is not known for arbitrary $m$. However, as follows from Eqs.~(\ref{eq:tauc},\ref{error_tau}), in the regime $m \gg t_P^{-\beta_d}$ ($\tau_m\gg t_P^{1-\beta_d}$), the proper time of events from $L^+_m$ is $\tau_m+O(t_P^{1-\beta_d})$. In this regime, we expect $\mbox{P}_d^+(\tau)$ to be a function sharply peaked at $\tau_m$, with fluctuations of the order $O(t_P^{1-\beta_d})$, so that the integral over proper times can be approximated as
\begin{equation}
\int_0^\infty \mbox{P}_d^+(\tau) \tau^d d\tau \propto  \tau_m^d t_P^{1-\beta_d},
\end{equation}
and that the number of events in $L^+_m$ becomes
\begin{equation}
N^+_m=\sigma_d t_P^{-(d+\beta_d)} \tau_m^d {}_2F_1\left(\frac{1}{2},\frac{d}{2},\frac{d+2}{2},-\sinh^2{\chi_c} \right)\sinh^d{\chi_c},
\end{equation}
where
\begin{equation}
\sinh{\chi_c}=\sqrt{\cosh^2{\chi_c}-1}=\frac{l_c}{\tau_m}\sqrt{1+\frac{l_c^2}{4\tau_m^2}},
\end{equation}
${}_2F_1()$ is the Gauss hypergeometric function, and $\sigma_d$ a constant that can be measured from numerical simulations.

Let us now set $l_c$ to be proportional to the proper time $\tau_m$, so that $\chi_c$ becomes a constant. This setting satisfies the requirements in Eq.~\eqref{eq:taum-lc-scaling}. With this setting, the number of nodes in $L_m^+$ scales as
\begin{equation}
N^+_m \sim t_P^{-(d+\beta_d)} \tau_m^d.
\end{equation}
We note that $N^+_m \gg 1$ in the regime $t_P^{1-\beta_d} \ll \tau_m \ll 1$.

\section{Evaluation of $\langle t^{+2} \rangle$ and $\langle x^{+2} \rangle$}
\label{app:t+}
In hyperbolic coordinates, the temporal coordinate of an event in $L^+_m$ is $t^+=\tau^+ \cosh{\chi}$, where $\tau^+$ and $\chi$ are statistically independent random variables, the former in the range $(0,\infty)$ and the latter in $(0,\chi_c)$, where $\chi_c$ is given by Eq.~\eqref{eq:chic}. The second moment of $t^+$ can then be computed as
\begin{equation}
\langle t^{+2} \rangle=  \frac{\int_0^\infty \mbox{P}_d^+(\tau) \tau^{d+2} d\tau \int_0^{\chi_c}\cosh^2{\chi}\sinh^{d-1}{\chi} d\chi}{ \int_0^\infty \mbox{P}_d^+(\tau) \tau^d d\tau \int_0^{\chi_c}\sinh^{d-1}{\chi} d\chi}.
\end{equation}
In the regime $\tau_m \gg t_P^{1-\beta_d}$ we obtain
\begin{align}
\langle t^{+2} \rangle&=f_t\left(\chi_c\right)\tau_m^2,\text{ where}\\
f_t(\chi_c)&=\frac{{}_2F_1\left(-\frac{1}{2},\frac{d}{2},\frac{d+2}{2},-\sinh^2{\chi_c} \right)}{{}_2F_1\left(\frac{1}{2},\frac{d}{2},\frac{d+2}{2},-\sinh^2{\chi_c} \right)}.
\end{align}
We note that $f_t(\chi)>1$ for $\chi>0$. Using a similar calculation, it is easy to show that the variance of $t^+$ scales as $\mbox{Var}(t^+)\sim \langle t^{+2} \rangle$.

Finally, the second moment of individual space coordinates follows from the identity $t^{+2}=\tau^{+2}+r^{+2}$ and the symmetry of $\mathbb{M}^{d+1}$, so that
\begin{equation}
\langle x^{+2} \rangle=\frac{1}{d}\left( \langle t^{+2} \rangle-\tau_m^2 \right),
\end{equation}
which, if $\tau_m \gg t_P^{1-\beta_d}$, is
\begin{equation}
\langle x^{+2} \rangle=f_x(\chi_c)\tau_m^2 = \frac{1}{d}\left(f_t(\chi_c)-1\right)\tau_m^2.
\end{equation}

If $l_c\sim\tau_m$, then $\chi_c$ is a constant, and
\begin{equation}
\langle t^{+2} \rangle \sim \langle x^{+2} \rangle \sim \tau_m^2.
\end{equation}
By choosing the constant $C>0$ in $l_c=C\tau_m$, we can fix the constant $C_x$ in
\begin{equation}
\langle x^{+2} \rangle = C_x \tau_m^2
\end{equation}
to any positive value (e.g.,~$1$), in which case $\langle t^{+2} \rangle$ becomes
\begin{align}
\langle t^{+2} \rangle &= C_t \tau_m^2, \text{ where}\\
C_t &= C_xd+1.
\end{align}

%\bibliographystyle{revtex}
%\bibliography{bib}

%apsrev4-2.bst 2019-01-14 (MD) hand-edited version of apsrev4-1.bst
%Control: key (0)
%Control: author (8) initials jnrlst
%Control: editor formatted (1) identically to author
%Control: production of article title (0) allowed
%Control: page (0) single
%Control: year (1) truncated
%Control: production of eprint (0) enabled
%

\end{document}